\documentclass[trackchanges,twocolumn]{aastex701}
\usepackage{amsmath}
\usepackage{upgreek}
\newcommand{\um}{\textmu m\,}
\providecommand{\e}[1]{\ensuremath{\times 10^{#1}}}

\begin{document}

\title{A carbon-rich atmosphere on a windy pulsar planet}


\author[orcid=0000-0002-0659-1783]{Michael Zhang}
\altaffiliation{51 Pegasi b Fellow}
\affiliation{University of Chicago, Department of Astronomy \& Astrophysics}
\email[show]{mzzhang2014@gmail.com}  

\author[orcid=0000-0003-1285-8170]{Maya~Beleznay} 
\affiliation{Stanford University, KIPAC}
\email{mayabel@stanford.edu}

\author[orcid=0000-0003-2630-8073]{Timothy D. Brandt}
\affiliation{Space Telescope Science Institute}
\email{tbrandt@stsci.edu}

\author[orcid=0000-0001-6711-3286]{Roger W. Romani}
\affiliation{Stanford University, KIPAC}
\email{rwr@stanford.edu}

\author[0000-0002-8518-9601]{Peter Gao}
\affiliation{Carnegie Science}
\email{pgao@carnegiescience.edu}

\author[0000-0002-6980-052X]{Hayley Beltz}
\affiliation{University of Maryland, Department of Astronomy}
\email{hbeltz@umd.edu}

\author[0000-0003-3294-3081]{Matthew Bailes}
\altaffiliation{OzGrav: The ARC Centre of Excellence for Gravitational Wave Discovery}
\affiliation{Swinburne University of Technology, Centre for Astrophysics and Supercomputing}
\email{mbailes@swin.edu.au}

\author[0000-0001-8236-5553]{Matthew C. Nixon}
\altaffiliation{51 Pegasi b Fellow}
\affiliation{Arizona State University, School of Earth and Space Exploration}
\affiliation{University of Maryland, Department of Astronomy}
\email{mcnixon@umd.edu}

\author[0000-0003-4733-6532]{Jacob L. Bean}
\affiliation{University of Chicago, Department of Astronomy \& Astrophysics}
\email{jacobbean@uchicago.edu}

\author[0000-0002-9258-5311]{Thaddeus D. Komacek}
\affiliation{University of Oxford, Department of Physics (Atmospheric, Oceanic and Planetary Physics)}
\email{tkomacek@umd.edu}

\author[0000-0002-0508-857X]{Brandon P. Coy}
\affiliation{University of Chicago, Department of Geophysical Sciences}
\email{bpcoy@uchicago.edu}

\author[0000-0002-3263-2251]{Guangwei Fu}
\affiliation{Johns Hopkins University, Department of Physics and Astronomy}
\email{guangweifu@gmail.com}

\author[0000-0002-3263-2251]{Rafael Luque}
\altaffiliation{NHFP Sagan Fellow}
\affiliation{University of Chicago, Department of Astronomy \& Astrophysics}
\email{rluque@uchicago.edu}

\author[0000-0002-2035-4688]{Daniel J. Reardon}
\altaffiliation{OzGrav: The ARC Centre of Excellence for Gravitational Wave Discovery}
\affiliation{Swinburne University of Technology, Centre for Astrophysics and Supercomputing}
\email{dreardon@swin.edu.au}

\author[0000-0003-3265-2866]{Emma Carli}
\altaffiliation{OzGrav: The ARC Centre of Excellence for Gravitational Wave Discovery}
\affiliation{Swinburne University of Technology, Centre for Astrophysics and Supercomputing}
\email{ecarli@swin.edu.au}

\author[0000-0002-7285-6348]{Ryan M. Shannon}
\altaffiliation{OzGrav: The ARC Centre of Excellence for Gravitational Wave Discovery}
\affiliation{Swinburne University of Technology, Centre for Astrophysics and Supercomputing}
\email{rshannon@swin.edu.au}

\author[0000-0002-9843-4354]{Jonathan J. Fortney}
\affiliation{University of California Santa Cruz, Astronomy \& Astrophysics Department}
\email{jfortney@ucsc.edu}

\author[0000-0002-4487-5533]{Anjali A.A. Piette}
\affiliation{University of Birmingham, School of Physics \& Astronomy}
\email{a.a.a.piette@bham.ac.uk}

\author[0000-0002-2666-728X]{M. Coleman Miller}
\affiliation{University of Maryland, Department of Astronomy}
\email{mcmiller@umd.edu}

\author[0000-0002-0875-8401]{Jean-Michel Desert}
\affiliation{University of Amsterdam, Anton Pannekoek Institute for Astronomy}
\email{J.M.L.B.Desert@uva.nl}

\begin{abstract}
A handful of enigmatic Jupiter-mass objects have been discovered orbiting pulsars.  One such object, PSR\,J2322-2650b, uniquely resembles a hot Jupiter exoplanet due to its minimum density of 1.8\,g\,cm$^{-3}$ and its $\sim$1,900\,K equilibrium temperature.  We use JWST to observe PSR\,J2322-2650b's emission spectrum across an entire orbit.  In stark contrast to every known exoplanet orbiting a main-sequence star, we find an atmosphere rich in molecular carbon (C$_3$, C$_2$) with strong westward winds.  Our observations open up new exoplanetary chemical (ultra-high C/O and C/N ratios of $>100$ and $>10,000$, respectively) and dynamical regimes (ultra-fast rotation with external irradiation) to observational study.  The extreme carbon enrichment poses a severe challenge to the current understanding of ``black widow'' companions, which were expected to consist of a wider range of elements due to their origins as stripped stellar cores.
\end{abstract}


\section{Introduction} 
In ``black widow'' systems, a millisecond pulsar is orbited closely ($P_{\rm orb} < 1$ d) by a low mass ($<$0.1 $M_\odot$), often degenerate companion. 
They are so named because in a prior low-mass X-ray binary phase the pulsar was spun up by accreting mass from the companion, while at present the companions are being evaporated by the pulsar. There are $\sim$50 known black widow systems \citep{koljonen_2025}, several which have estimated companion masses below 10 $M_J$.
Of the handful with short ($\lesssim$ few day) orbital periods, only one has a minimum density similar to that of gas giants orbiting main sequence stars: PSR J2322-2650b \citep{spiewak_2017,shamohammadi_2024}, with a minimum mass of 0.8 $M_J$ and minimum density of 1.8 g cm$^{-3}$.  The pulsar's unusually low spindown luminosity of $2 \times 10^{32}$ erg\,s$^{-1}$ --- arising from its unusually low magnetic field --- would give the companion an equilibrium temperature of 1300 K if isotropically radiated and fully converted to heat.  However, the gamma rays that dominate a millisecond pulsar's energy output and are responsible for heating the companion are typically beamed toward the equator \citep[e.g.][]{philippov_2020}, making an accurate prediction of companion temperature difficult.

The atmospheres of a few of PSR J2322-2650b's denser and more massive black widow cousins have been studied from the ground (e.g. \citealt{draghis_2019,kandel_2020,dhillon_2022}). \cite{kandel_2023} summarizes ten sets of previously published photometric phase curves of black widow companions, with masses ranging from 12 to 67 $M_J$ and densities ranging from 20 to 40 g/cm$^3$.  All are extremely hot --- nine have \textit{night side} temperatures above 2200 K, while J1311-3430 has no detectable night side emission, but its day side is $>$10,000 K.  These studies have measured the atmospheric circulation, Roche lobe fill factors, and inclinations of these extreme objects.  They have even provided tentative estimates of pulsar masses and put constraints on the neutron star equation of state.  Spectra have been taken of several black widow companions (e.g. \citealt{van_kerkwijk_2011,romani_2015,romani_2016,kennedy_2022,romani_2022,simpson_2025}) and are generally stellar, indicating roughly Solar composition.  However, J1311$-$3430 and J1653$-$0158 both have $10-15M_J$ companions with H-free surface spectra; these represent a black-widow sub-class, the `Tidarrens', which are likely descendants of ultra-compact low mass X-ray binaries with evolved companions \citep{romani_2015}.  \cite{guo_2022} showed with MESA simulations that these ultra-light companions, including PSR J1719-1438, J2322-2650, and J1311-3430, could have formed by the Roche lobe overflow and photoevaporation of helium stars.

\begin{figure}[ht]
    \includegraphics[width=0.5\textwidth]{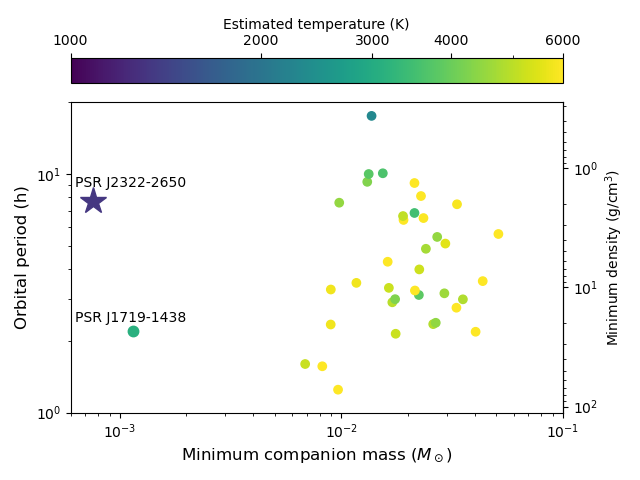}
    \caption{PSR J2322-2650 compared to the black widows from SpiderCat \citep{koljonen_2025} and to PSR J1719-1438 \citep{bailes_2011}, showing it is in a distinct area of parameter space.  PSR J1719-1438, though similar in mass, has a much higher mean density of 21 g/cm$^3$; its bulk composition is likely very different.  The temperature of all companions was very roughly estimated by assuming the spindown luminosity is isotropically radiated and all of the energy hitting the companion is converted to heat.  The minimum density was derived from the orbital period via $\rho_{\rm min} = 3\pi/(0.46^3 G P^2)$, combining Paczy{\'n}ski's approximation for the Roche lobe ($R_L = 0.46^3 q^{1/3} a$; \citealt{paczynski_1971}) with Kepler's third law.}
    \label{fig:black_widows}
\end{figure}

PSR J2322-2650b is different from other ultra-light pulsar companions, being the only pulsar companion with a mass, density, and temperature similar to that of hot Jupiters (Figure \ref{fig:black_widows}).  The atmosphere of such an object has never been observed.  It is however very easily observable with JWST: at infrared wavelengths, the pulsar is undetectable, giving us the unprecedented opportunity to obtain exquisite spectra of an externally irradiated planetary-mass object.  In this paper, we present JWST spectra of this exotic, gamma-ray-heated ``exoplanet'', unveiling a bizarre atmosphere that raises more questions than it answers.

\section{Observations}
On 8 Nov 2024, we used NIRSpec/PRISM on JWST to observe PSR\,J2322-2650b's 0.6--5.3 \um spectroscopic phase curve.  These observations started shortly before inferior conjunction (the ``night side'') and continued until shortly after the following inferior conjunction, covering the whole 7.8 h orbit.  Two days later, we used NIRSpec/G235H (1.7--3.1 \um) to take a 2 h-long higher-resolution spectral sequence bracketing superior conjunction (the ``day side'').  This data set spans an orbital phase range of 0.25, allowing us to measure the radial velocity of the planet as a function of time.  Both the PRISM and G235H observations are taken with the 0.2$^{\prime\prime}$ fixed slit.   All the JWST data used in this paper can be found in MAST: \dataset[10.17909/shv0-yq03]{http://dx.doi.org/10.17909/shv0-yq03}.  We develop a custom pipeline to reduce the data, significantly reducing the scatter compared to JWST standard pipeline processing (see Appendix \ref{sec:jwst_pipeline}). 

\begin{figure*}[ht]
    \includegraphics[width=\textwidth]{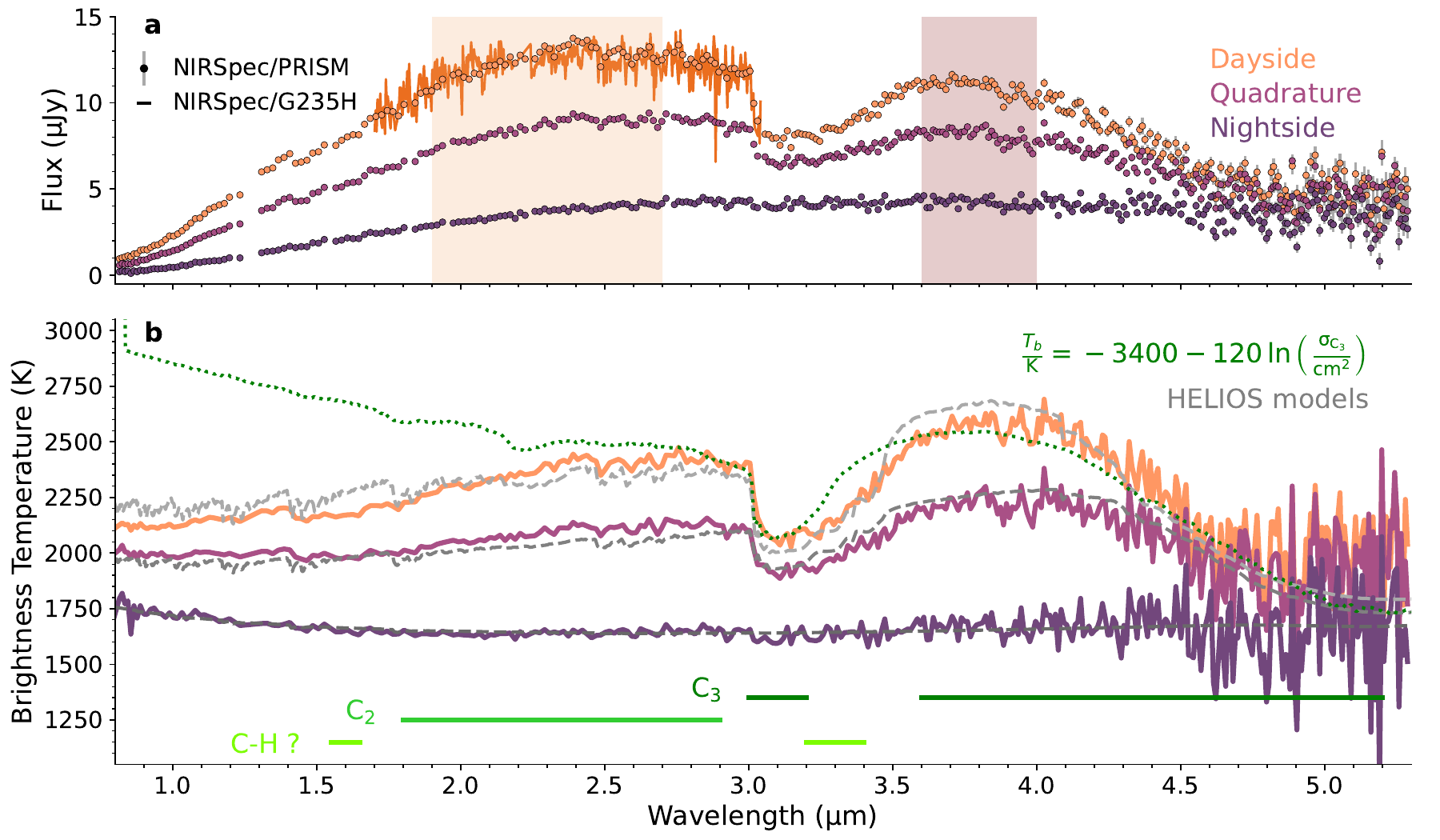}
    \caption{The observed emission flux and brightness temperature of PSR J2322-2650b. \textbf{a}, observed PRISM spectrum of the dayside, the average of the two quadratures, and nightside at native resolution.  The time-averaged G235H spectrum is also plotted, binned in wavelength by a factor of 16. The two shaded wavelength regions are those of the light curves in Fig. \ref{fig:wlc}.  \textbf{b}, PRISM spectra expressed in brightness temperatures by assuming R/D = (1.1 $R_J$) / (630 pc).  We plot HELIOS radiative-convective equilibrium forward models in grey.  A linear function of the log of C$_3$'s absorptions cross sections is plotted in green, showing that C$_3$ absorption explains the sudden dip at 3.014 um, the recovery at 4 um, and the slow decline toward the red edge.}
    \label{fig:spectra}
\end{figure*}

We show in Figure \ref{fig:spectra} the average PRISM spectrum on the dayside, quadrature, and nightside (defined as phases $0.5 \pm 0.085$, $0.25 \pm 0.05$ or $0.75 \pm 0.05$, and $0.00 \pm 0.05$ or $1.00 \pm 0.05$, respectively).  The spectra are shown both in $\upmu$Jy and in brightness temperature, calculated with a fiducial radius of $1.1 R_J$ and an updated radio pulsar timing parallax distance of 630 pc (see Appendix \ref{sec:radio_analysis}).  While the nightside spectrum is featureless and consistent with a near-isothermal temperature profile or a thick grey dust/cloud deck, the dayside spectrum has clear absorption features.  We can identify the molecules giving rise to these features by comparing the brightness temperature spectrum against $-\ln{(\sigma(\lambda))}$, where $\sigma(\lambda)$ is the absorption cross section of a candidate molecule.  After an complete search of the DACE \footnote{\url{https://dace.unige.ch/opacity/}} molecular database and an extensive search of the ExoMol \citep{tennyson_2018} molecular database, we conclude that C$_3$ and C$_2$ are firmly detected.  As shown in Figure \ref{fig:spectra}, the cross section of $^{12}$C$_3$ \cite{lynas-gray_2024} rises suddenly redward of 3.014 \um, matching the sudden flux drop seen in the data.  (The isotopologues $^{12}$C$^{13}$C$^{12}$C and $^{12}$C$^{12}$C$^{13}$C have opacity cliffs at slightly different wavelengths and are inconsistent with the data, but their line lists are less reliable because they are not matched to experiment.)  C$_3$ also has an opacity minimum at 3.9 \um, matching the maximum in the emission spectrum, and an opacity maximum at 5.1 \um matching the emission plateau.  The sawtooth pattern at 2.45--2.85 \um matches the series of band-heads in the opacity of C$_2$ \citep{yurchenko_2018}.

C$_3$ and C$_2$ are necessary but not sufficient to explain the PRISM spectrum.  First, the absorption feature at 3.0--3.6 \um is wider than the C$_3$ opacity peak, requiring another source of opacity around 3.4 \um.  Similarly, while a C$_2$ opacity peak around 1.4 \um could explain the sudden drop in flux at that wavelength, the absorption feature extends much redder than the opacity peak, suggesting that there is missing opacity around 1.6 \um.  We confirm that isotopologues of C$_3$ and C$_2$ cannot explain the missing opacity at either wavelength, and tentatively identify both absorption features as arising from the C-H bond.  The fundamental stretching mode of this bond is at 3.4 \um, resulting in the ubiquitous presence of absorption at this wavelength among organic molecules [Table 1.2; \citep{tammer_2004}].  The first overtone of the 3.4 \um stretch mode is at half the wavelength, or 1.7 \um.  Due to the similarity of absorption features across large classes of organic molecules, the exact molecule(s) giving rise to the C-H absorption is difficult to determine.  We consider the identification of the features as C-H absorption to be tentative because we detect no absorption feature around 2.3 \um, whereas such a feature is common in hydrocarbons.

\begin{figure*}[ht]
    \includegraphics[width=\textwidth]{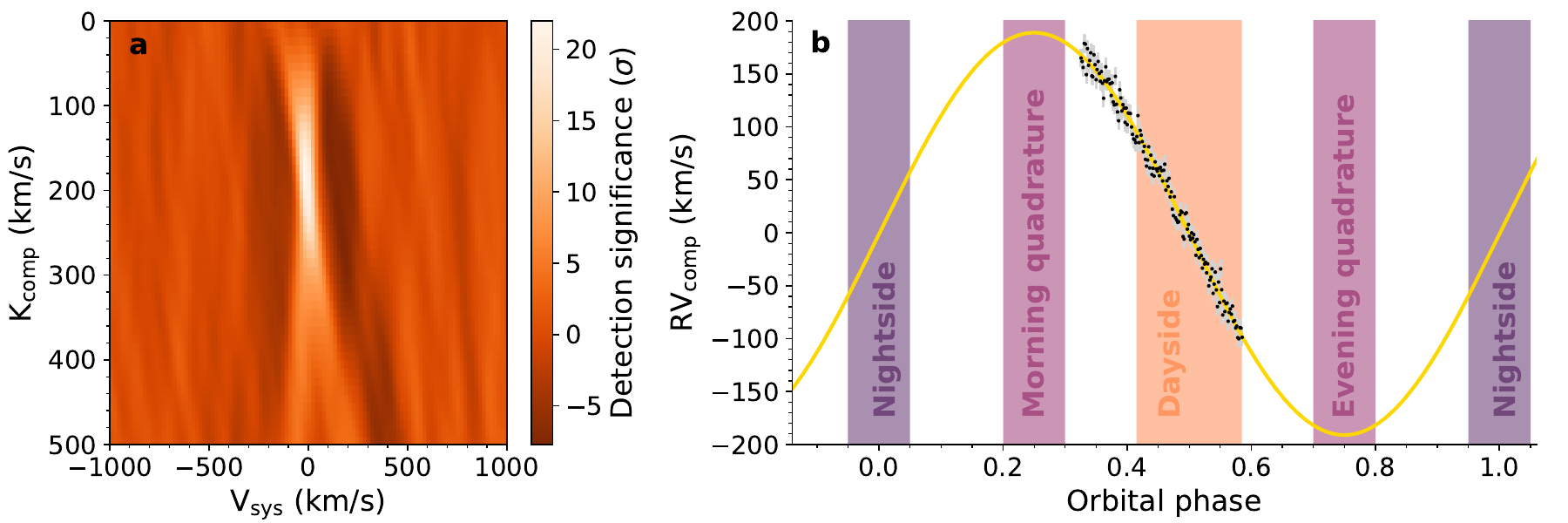}
    \caption{C$_2$ detection significance and radial velocities obtained from the JWST NIRSpec/G235H data.  \textbf{(a)} The detection significance from cross correlation of C$_2$ in the G235H data, as a function of projected orbital velocity and barycentric radial velocity.  C$_2$ is detected at 21$\sigma$ at V$_{\rm sys} \approx 0$ and $K_{\rm comp}=190$ km/s.  \textbf{(b)} In black, the companion radial velocity measured from each integration of the G235H observations using a data-derived template; in yellow, a sinusoidal fit to the data.  The vertical offset is arbitrary.  In the colored bars, we show the phases we define to be the nightside, quadrature, and dayside for the purposes of calculating the average spectra in Fig. \ref{fig:spectra}.}
    \label{fig:high_res}
\end{figure*}

While the sawtooth feature in the PRISM spectrum strongly hints at the existence of C$_2$, the higher-resolution G235H spectra conclusively demonstrate its existence.  Using techniques from high-resolution cross correlation spectroscopy \citep{snellen_2025}, we cross correlate the smoothed logarithm of C$_2$ opacities with each of the 151 integrations in the grating observation (see Appendix \ref{sec:cross_corr}).  Plotting the cross correlation function (CCF) with respect to time and blueshift reveals the radial velocity track of the planet.  Given an assumed projected orbital velocity $K_p$ and barycenter velocity $v_{\rm sys}$, we can shift the CCFs into the planetary frame and sum them along the time axis to obtain a C$_2$ detection significance.  Maximizing this significance over the $K_p$-$v_{\rm sys}$ plane, as shown in the left panel of Figure \ref{fig:high_res}, yields a 21$\sigma$ detection of C$_2$ at a barycentric velocity close to 0 and a projected orbital velocity $K_p \sim 200$ km/s.

\section{Composition constraints}
In exoplanetary atmospheres, carbon-bearing molecules such as CO, CH$_4$, and CO$_2$ are commonly detected, but molecular carbon has never been seen.  To quantify how depleted the atmosphere is in non-carbon elements, we obtain robust model-independent constraints on molecular abundance ratios by constructing a simulated G235H spectrum from only the cross sections of C$_2$ and a non-C$_2$ molecule, adjusting the line contrast until C$_2$ is detected at the same significance as in the real data, and computing the detection significance of the other molecule.  As detailed in Appendix \ref{sec:cross_corr}, we perform this procedure with C$_2$/CO using the 2.30--2.47 \um range, where both molecules have substantial opacity, and find a 3$\sigma$ lower limit on C$_2$/CO of 0.17. We repeat the procedure with CN, which has higher opacity than CO over a larger wavelength range, finding C$_2$/CN $> 32$ at 3.2$\sigma$.  We additionally search for a wide variety of other molecules in the grating data via cross correlation--including CS, CH, NS, OCS, H$_2$O, and CH$_4$--but find no $>3\sigma$ detections.

Our inferred abundances suggest an atmosphere rich in carbon and relatively depleted in oxygen and nitrogen. Hydrogen is also heavily depleted, as otherwise the ample carbon atoms would bond with H to form hydrocarbons resulting in minuscule abundances of molecular carbon, as has been predicted for hot Jupiter exoplanets with high C/O ratios \citep{madhusudhan_2012}.  Some C-H features are still possible, however, as trace amounts of hydrogen can be produced via spallation by the pulsar wind \citep{hansen_1996}. On the other hand, this object is unlikely to be dominated by carbon in bulk due to its equation of state.  To test this, we calculate the mass-radius relationship of a pure carbon object using an interior structure model \citep{nixon_2021,nixon_2024} with the equation of state of \cite{kerley_2001} (see Appendix \ref{sec:interior_modelling}) and find that, at zero temperature, the radius of a 1.5--2.5 $M_J$ object (0.39 $R_J$) is much smaller than the near-Roche-lobe-filling radius our PRISM phase curve requires (see Appendix \ref{sec:prism_phase_curve}).  A central temperature of 500,000 K would be needed to inflate a carbon object to a Jupiter radius, but this is unrealistically hot, given (as we will later argue) its likely past as a white dwarf with efficient conductive heat transport.  If, however, we assume an object that is mostly helium, with 1\% C, then the radius increases to 0.92 $R_J$ for a 1.5 $M_J$ object with a photosphere temperature of 1500~K, using the helium equation of state from \citep{chabrier_2021} and an additive volume law.  We thus conclude that the planet's bulk composition is likely helium-dominated.

\subsection{Equilibrium chemistry}
\begin{figure*}[ht]
\centering
    \includegraphics[width=0.9\textwidth]{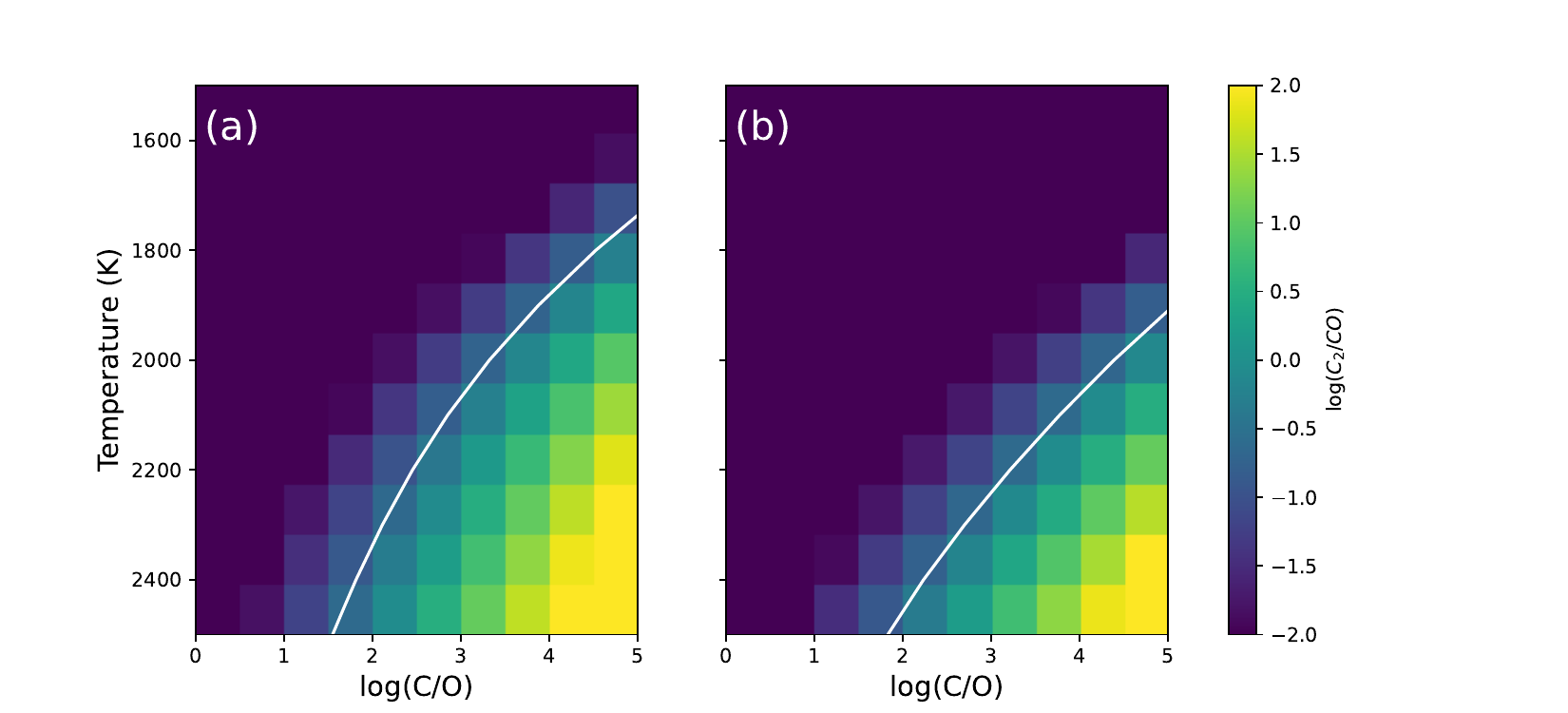}
    \caption{The relative abundance of C$_2$ to CO in chemical equilibrium.  \textbf{(a)} Abundance ratios for C/He = $10^{-2}$ and P=10 mbar, for an atmosphere with no other elements. \textbf{(b)} Ratios in an atmosphere with H/He=0.01 as well as solar values of N/He, P/He, S/He.  The solid white curves represent the 3$\sigma$ constraint from our cross correlation test (C$_2$/CO $>$ 0.17).}
    \label{fig:C2_to_CO}
\end{figure*}

We can test whether a helium-dominated atmosphere enriched in carbon can reproduce our observations by performing equilibrium chemistry calculations with FastChem \citep{stock_2018,stock_2022}.  Figure \ref{fig:C2_to_CO} plots the ratio of C$_2$/CO as a function of temperature and atomic C/O, assuming C/He=0.01 and either no other elements, or assuming H/He=0.01 as well as solar N/He, P/He, and S/He.  In either case CO is highly favored over C$_2$ until C/O $\gg$ 1, and far more favored at low temperatures.  Even at 2300 K, the hottest point in our ICARUS model (see below), our constraint of C$_2$/CO$>$0.17 requires C/O $>$ 100.  At the average dayside brightness temperature of 2000 K, C$_2$/CO$>$0.17 would require C/O$>2000$ (pure He and C scenario) or C/O$>20,000$ (many-elements scenario).  We generate multidimensional abundance grids to explore a wide range of metallicities, C/O and C/N ratios, temperatures, and pressures, finding that our observational constraints on C$_2$/CO and C$_2$/CN imply C/O $>$ 100 and C/N $>$10,000 over a wide range of conditions (see Appendix \ref{sec:eq_chem_grid}).

In our high C/O calculations with only C, O, and He, C$_3$ is the most abundant molecule, consistent with our observations; at C/O $\sim$ 1, CO is the most abundant molecule and C$_3$ has negligible (parts per billion or less) abundances.  In our calculations which include hydrogen and other elements, C$_3$H and C$_2$H are comparably abundant to C$_3$ and C$_2$ over a wide range of temperatures and C/O ratios.  The C-H bonds on these molecules may explain the opacity we see at 3.4 and 1.6 \um, but since there are no published line lists for these molecules, we are unable to include them in our modeling.  Colder carbon-rich compositions (T$<$1000 K, C/O $\gg$ 1) are predicted to have abundant quantities of larger carbon molecules, such as C$_4$ and C$_5$, which also have no published line lists. To compare these results to data, we generate forward models using HELIOS \citep{malik_2017,malik_2019} (see Appendix \ref{sec:helios}), assuming 1D radiative-convective equilibrium in a bottom-heated helium-dominated atmosphere with an extended graphite dust/cloud layer and $~\sim$ppt abundances of C$_3$, C$_2$, and C$_2$H$_4$ (a stand-in for a generic molecule with C-H bonds). These are shown in Figure \ref{fig:spectra}, which demonstrates that this simple prescription reproduces the main spectral features at all phases, with the featureless night side explained by invoking more dust absorption. 

\section{Radial velocities}
Refining the value of the projected orbital velocity $K_p$ allows us to pin down the inclination of the companion orbit and the masses of the companion and pulsar, as well as assess the validity of our fiducial companion radius. We use the G235H spectra to monitor the radial velocity variations \citep{sing_2024}.  We generate a template spectrum by shifting every spectrum into the planetary frame using our preliminary $K_p$ value ($\sim$200 km/s) and taking the time-wise average.  We then fit this template to each individual spectrum to find the radial velocity shift as a function of time, as shown in Figure \ref{fig:high_res} (right).  Finally, we fit the sinusoid $v(\phi) = v_0 + K_p \sin{(2\pi\phi)}$ to the velocities.  We obtain $K_p = 190 \pm 2$ km/s.  This value represents the center of light orbital velocity $K_{CoL}$, which is lower than the planet's center-of-mass orbital velocity $K_{CoM}$ because the bright, irradiated side of the planet is closer to the pulsar.  The ratio $K_{\rm cor} = K_{CoM}/K_{CoL}$ depends on the heating pattern, temperature dependence of the lines dominating the radial velocity, the Roche lobe fill factor, and inclination.  For our system with its mass ratio $q \approx 1000$, the maximum possible value of $K_{\rm cor}$ is $1/(1 - (3q)^{-1/3}) = 1.07$, which reduces to 1.04 for a more realistic $T_0 \cos^{1/4}(\theta)$ temperature distribution \citep{van_kerkwijk_2011}.

For a binary system in a circular orbit where the primary is much more massive than the secondary, $K_{\rm CoM} = K_{CoL}K_{\rm cor} = (\frac{2\pi G M_*}{P})^{1/3} \sin{i}$.  Assuming the pulsar mass $M_p$ can range from the canonical $1.4 M_\odot$ to an unusually massive $2.4 M_\odot$ (c.f. Fig 7 of \citealt{lattimer_2012}), we infer an inclination range of $34.7-28.4^\circ$ (for $K_{\rm cor}=1.04$). This low $i$, combined with the fact that pulsar $\gamma$-rays are beamed toward the spin equator 
explains a number of system features: the lack of a {\it Fermi} $\gamma$-ray detection \citep{ballet_2023}, the high companion temperature compared to the pulsar spindown power, and the lack of radio eclipses from a companion outflow.  This range of inclinations implies a planet mass of $1.4-2.4 M_J$, Roche lobe radius $R_L = 0.462a(m_p / m_*)^{1/3} = 0.462(\frac{GM_pP^2}{4\pi^2})^{1/3}$  of $0.99-1.18 R_J$, and density of 1.8 g/cm$^3$.  All plausible pulsar masses therefore imply a roughly Jupiter-sized and Jupiter-mass companion on a $\sim30^\circ$ inclination orbit.

\section{PRISM phase curve}
\label{sec:prism_phase_curve}

\begin{figure*}[t]
    \centering
    \includegraphics[width=1.0\textwidth]{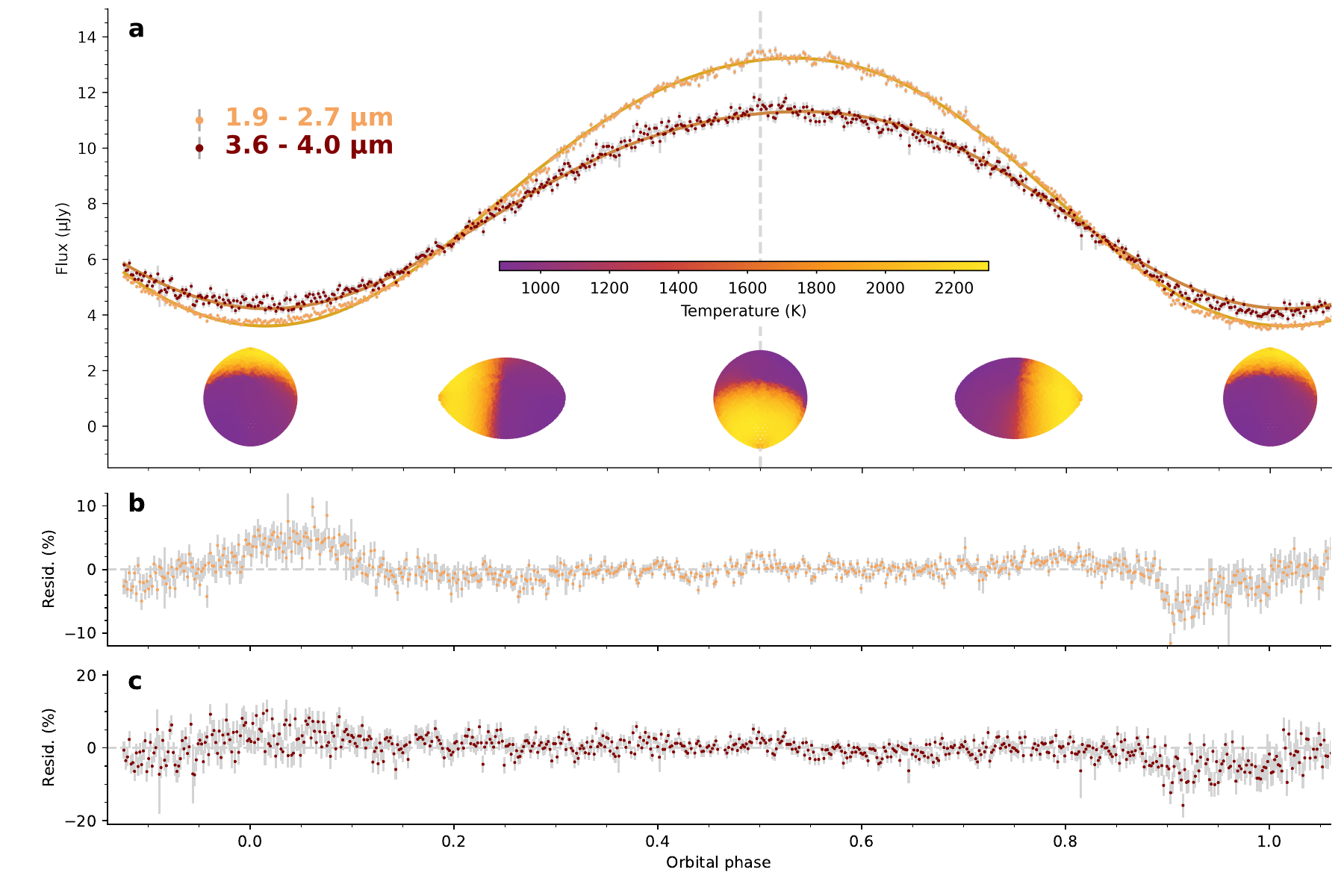}
    \caption{The companion light curve in two continuum bands interpreted with a pulsar direct heating model. \textbf{(a)} The models are shown by the curves and the data are shown by the points. The model does not capture the strong orbit-to-orbit variations at minimum ($\phi \sim 0$ and $\phi \sim 1$), but provides a good description of the flux from the heated face and constrains the binary parameters (see Appendix \ref{sec:prism_phase_curve}). Schematics of the Earth-view of the heated companion's surface temperature are shown below the light curve. \textbf{(b)},\textbf{(c)} The residuals of the models compared to their corresponding light curves.}
    \label{fig:wlc}
\end{figure*}

Next, we model the PRISM phase curve.  The C$_3$ absorption depth varies dramatically with orbital phase, complicating modeling of the heated companion surface. However, at wavelengths of relatively low opacity, a basic thermal model can be used to constrain the binary properties. Figure \ref{fig:wlc} shows the light curves in two wavelength ranges between the strongest absorption bands. The flux varies by $\sim 3\times$ from day to night, a modest amplitude compared to other black widow companions, which can show modulations of $30-100\times$ \citep{draghis_2019}.  The light curves exhibit nightside variability--the second minimum is 10\% lower than the first, possibly due to flares \citep{kandel_2020}.  We model the light curves in the two opacity windows by assuming Planckian emergent spectra and direct pulsar heating, using a version of the ICARUS heated binary light curve code \citep{breton_2012} (see Appendix \ref{sec:prism_phase_curve}). 
All fits require a near-filled Roche lobe.  Fitting the full light curves, we find $i\sim 31^\circ$ with photosphere temperatures ranging from 900 K at the coldest points on the night side, to 2300 K at the hottest points of the day side.  We infer a distance consistent with the radio parallax measurement and a westward wind.  Given the closeness of the inferred inclination to the values we obtained previously by assuming reasonable pulsar masses, it is not surprising that we now infer a pulsar mass ($\sim 2.0 M_\odot$ with 0.1 $M_\odot$ statistical error and $0.5 M_\odot$ systematic error) consistent with astrophysical expectations (see Appendix \ref{sec:prism_phase_curve}).  The inferred isotropic heating luminosity of $9 \times 10^{32}$ erg\,s$^{-1}$ is $5/I_{45}$ times the pulsar's spindown luminosity where $10^{45} I_{45}$ g cm$^2$ is the pulsar's moment of inertia.  The apparent over-unity heating efficiency is an artifact of the pulsar's gamma ray emission being beamed along the spin equator (and hence, toward the companion), possibly in combination with a large moment of inertia arising from a high pulsar mass.

The flux maximum in the phase curve occurs 12$^\circ$ after phase 0.5, indicating a westward offset that is suggestive of strong winds blowing opposite the planet's rotation direction.  A westward thermal phase offset is seen in some black widow companions \citep{kandel_2020} and rare among hot Jupiters orbiting main sequence stars \cite{bell_2021}, having only been robustly observed for CoRoT-2b \cite{dang_2018}.  All of these objects have rotation periods equal to their orbital periods due to tidal locking, but PSR\,J2322-2650b's exceptionally short rotation period of 0.32 d places it in a different dynamical regime from almost all hot Jupiters.  \cite{tan_2020,Lee:2020aa} demonstrated with 3D general circulation models (GCMs) that, at very short rotation periods of $\lesssim 10$ hours, the equatorial eastward jet ubiquitous on slower-rotating hot Jupiters narrows while off-equatorial westward zonal winds become increasingly distinctive, leading to a westward phase offset.  Because we do not expect the dynamics on PSR\,J2322-2650b to be identical to \cite{tan_2020} due to the very different form of irradiation, we run a planet-specific atmospheric dynamics model with the MITgcm (see Appendix \ref{sec:gcms}).  Our temperature maps are qualitatively similar to theirs; in particular, we also obtain a westward phase offset.  Our data are therefore observational evidence of the dynamical regime that \cite{tan_2020} predicted, a regime rarely probed by hot Jupiters orbiting main sequence stars.

Better line lists and improvements in the modeling of planetary atmospheres will allow more accurate computation of inclination and pulsar mass through modeling of the spectroscopic phase curve.  This could be particularly interesting for PSR\,J2322$-$2650 because the pulsar's low intrinsic spindown ${\dot P}$ implies an exceptionally low neutron star dipole field $2.5 \times 10^7 I_{45}$\,G.  This is relevant since the reduced magnetic field of millisecond pulsars is often attributed to accretion during the spin-up process \citep{1974SvA....18..217B,taam_1986,1990Natur.347..741R,2017JApA...38...48M}; with such a low field the pulsar may have accreted a large amount of matter, achieving a larger moment of inertia and a high mass. Thus precise measurement of the masses in this system could help constrain $M_{\rm TOV}$, the Tolman-Oppenheimer-Volkoff mass, which sets the maximum mass for a slowly rotating neutron star and constrains the dense matter equation of state.

\section{Conclusion}
Our findings pose a challenge to the current understanding of black widow formation, in which a pulsar strips the outer layers of its stellar companion by a combination of Roche lobe overflow and photoevaporation \citep{benvenuto_2012,guo_2022}.  This mechanism can produce a helium-dominated Jupiter-mass object if stripping occurs before core helium burning, during the main sequence or red-giant branch.  In this picture these black widows are the descendants of ultra-compact X-ray binaries (UCXB) with He-star donors and the core C/N/O ratios can be sensitive to the evolutionary state of the donor at the start of Roche lobe overflow (Figure 4 of \cite{nelemans_2010}).  How this process produces a C/O ratio greater than 100 or C/N ratio greater than 10,000 is difficult to explain.  Although \cite{nelemans_2010} predicts extreme N depletion in one scenario (their Figure 4, bottom middle), O becomes significantly enhanced relative to C in this scenario.  In the other two scenarios (corresponding to different initial helium star masses), C is enhanced relative to O by a factor of $\sim$15, but it is only enhanced relative to N by a factor of $\sim$several.  Although extreme photospheric ratios are possible via gravitational settling for the much higher gravity white dwarfs (the immediate progenitors of black widow companions in the standard formation scenario), gravitational settling is not expected to differentiate elements once the companion becomes a planetary mass object with a molecular atmosphere. 

Other mechanisms of carbon enrichment pose their own challenges.  For example, the helium dominated, hydrogen poor, and carbon enriched composition of PSR J2322-2650b is reminiscent of R Coronae Borealis stars and dustless hydrogen-deficient carbon stars, which are believed to result from mergers of He and CO white dwarfs \citep{webbink_1984,saio_2002}.  However, the rarity of RCB/dLHdC stars and the necessity of invoking three stars makes them unlikely progenitors, and the measured C/O and C/N ratios of these stars (12--81 and 4--130 respectively; \cite{mehla_2025}) are still lower than the very high minimum values we obtain.  The triple alpha process, which fuses helium into carbon, creates AGB ``carbon stars'' with C/O ratios of up to $\sim$several \citep{abia_2008}, but not the C/O $>$ 100 inferred from our data.  Carbon stars have a dusty outflow rich in carbon grains \citep{di_criscienzo_2013}, providing a further source of carbon enrichment, but further modelling is required to explain how this carbon ultimately ends up in a Jupiter-mass object.  More spectroscopic observations of ultra-low-mass black widow companions (``Tidarrens'') are needed to determine whether PSR J2322-2650b's composition is unusual or representative of the class.  We particularly encourage observations of PSR J1719-1438, which has a similar mass but a far higher minimum density of 21 g/cm$^3$, leading \cite{bailes_2011} to suggest it is an ultralow-mass carbon white dwarf.

\section{Data access}
A Jupyter notebook is available to reproduce the main text figures (\url{https://doi.org/10.5281/zenodo.17400582}). The repository also includes the spectra plotted in Figure 2 (top) in ASCII format.

\begin{acknowledgments}
The authors thank Gerald Kerley for providing the carbon equation of state used for interior structure models. TDK acknowledges Jayne Birkby, Luke Parker, and Stephen Smartt for insightful discussions.  MZ thanks Qiao Xue, Keeyoon Sung, Sergey Yurchenko, John Tonry, and Jeremy Goodman for insightful feedback. MB thanks Jaikhomba Singha and Andrea Possenti for comments on the pulsar timing section, and Sarah Buchner for observing. PG acknowledges Shazrene Mohamed and Mikako Matsuura for enlightening discussions.

These observations are associated with JWST GO \#5263 and the work is funded by the associated grant.  The MeerKAT telescope is operated by the South African Radio Astronomy Observatory, which is a facility of the National Research Foundation, an agency of the Department of Science and Innovation. MZ acknowledges support from the 51 Pegasi b Fellowship funded by the Heising-Simons Foundation.  Part of this work was undertaken as part of the Australian Research Council Centre of Excellence for Gravitational Wave Discovery (project number CE230100016).  Support for this work was provided by NASA through the NASA Hubble Fellowship grant HST-HF2-51559.001-A awarded by the Space Telescope Science Institute, which is operated by the Association of Universities for Research in Astronomy, Inc., for NASA, under contract NAS5-26555.
\end{acknowledgments}

\begin{contribution}
M.Z. proposed the observations, led the project, identified C$_3$ and C$_2$, and led the writing of the paper.
M. Beleznay fit the continuum light curves with ICARUS to obtain temperature maps and constrain binary parameters.
T.D.B. performed the data analysis using primarily custom software.
R.W.R. gave advice on the pulsar physics, proposed a plausible formation mechanism, and helped with light curve fitting.
P.G. searched opacity databases to identify spectral features and ran atmospheric forward models to fit the spectrum.
H.B. ran, post-processed, and interpreted GCMs with a bespoke heating scheme.
M. Bailes is the PI of the MeerTime Large Survey Project, oversees the pulsar portal/pipeline, and created the standard profile, arrival times and initial model fit.
M.N. performed the interior structure calculations.
J.L.B. helped conceive the project and gave advice through all stages.
T.D.K. helped with GCM development and interpretation and writing the manuscript.
B.P.C. suggested the identification of C-H spectral features.
G.F. performed an independent data reduction.
R.L. helped with the figures and with data interpretation.
D.J.R. conducted pulsar timing analysis to determine system parameters.
E.C. ran and maintained the pipeline producing the MeerKAT pulsar timing data.
R.M.S. is the PI of the MeerKAT PTA timing project, and conducted pulsar timing analysis to determine system parameters.
J.J.F. advised on interior structure and atmosphere modeling.
A.A.A.P. performed a blackbody retrieval and gave comments on the manuscript.
M.C.M. edited the paper and contributed to understanding the pulsar heating.
J-M.D. contributed to the proposal.
\end{contribution}

%
\facilities{JWST(NIRSpec), MPTA(PTUSE)}

\software{astropy \citep{astropy:2013,astropy:2018,astropy:2022},  
          JWST Calibration Pipeline \citep{bushouse_2025},
          HELIOS \citep{malik_2017,malik_2019},
          HELIOS-K \citep{grimm_2021},
          FastChem \citep{stock_2022}
          }


\appendix

\section{Analysis of MeerKAT radio data}
\label{sec:radio_analysis}
The distance to the pulsar is important for the interpretation of this system in at least two ways: it determines the luminosity of the companion, and it allows us to derive the intrinsic spin-down rate $\dot P_{\rm i}$ due to the Shklovskii correction for pseudo-acceleration \citep{shklovskii_1970}.  The pulsar's spin-down luminosity and inferred magnetic field strength are both related to $\dot P_{\rm i}$.  The PSR~J2322$-$2650 discovery paper \cite{spiewak_2017} reported a low timing parallax distance of $230_{-50}^{+90}$\,pc, much smaller than their dispersion measure (DM) model distance of 760\,pc.  \cite{shamohammadi_2024}, using the more accurate MeerKAT radio telescope dataset, obtained a timing parallax of $1.3 \pm 0.2$ mas, from which derive a distance of 770$_{-100}^{+130}$\,pc.  Here, we improve the distance estimate by extending the MeerKAT dataset.

The 3.46\,ms pulsar J2322$-$2650 is observed regularly as part of the MeerKAT Pulsar Timing Array (MPTA; \citealt{miles_2023}) using the PTUSE pulsar instrument \citep{bailes_2020}. Observations span July 2019 to April 2025 and we obtain arrival times at 136 different epochs in 16 coherently dispersed frequency subbands spanning 775.0\,MHz centered at 1283.582\,MHz. A standard pulse profile is created by aligning all data with the optimal dispersion measure and with the phase-connected timing solution, removing the dispersion and averaging into 16 subbands before wavelet smoothing with the \textsc{psrchive} \citep{van_straten_2012} application, \textsc{psrsmooth}. Topocentric arrival times are measured with a standard Fourier-domain pulse profile cross-correlation technique, with uncertainties estimated with a Monte Carlo simulation. Arrival times with a signal-to-noise ratio of cross-correlation below 10 are removed and we perform a least-squares fit using \textsc{tempo2} \citep{Hobbs+06} for the pulsar spin, binary, and astrometric parameters on the 2625 remaining arrival times. The reduced $\chi^2$ is 1.0, the weighted RMS and the median residuals are 2.2\,$\upmu$s and 3.4\,$\upmu$s respectively, indicating a good model fit. 

\begin{figure}
    \centering
    \includegraphics[width=0.6\linewidth]{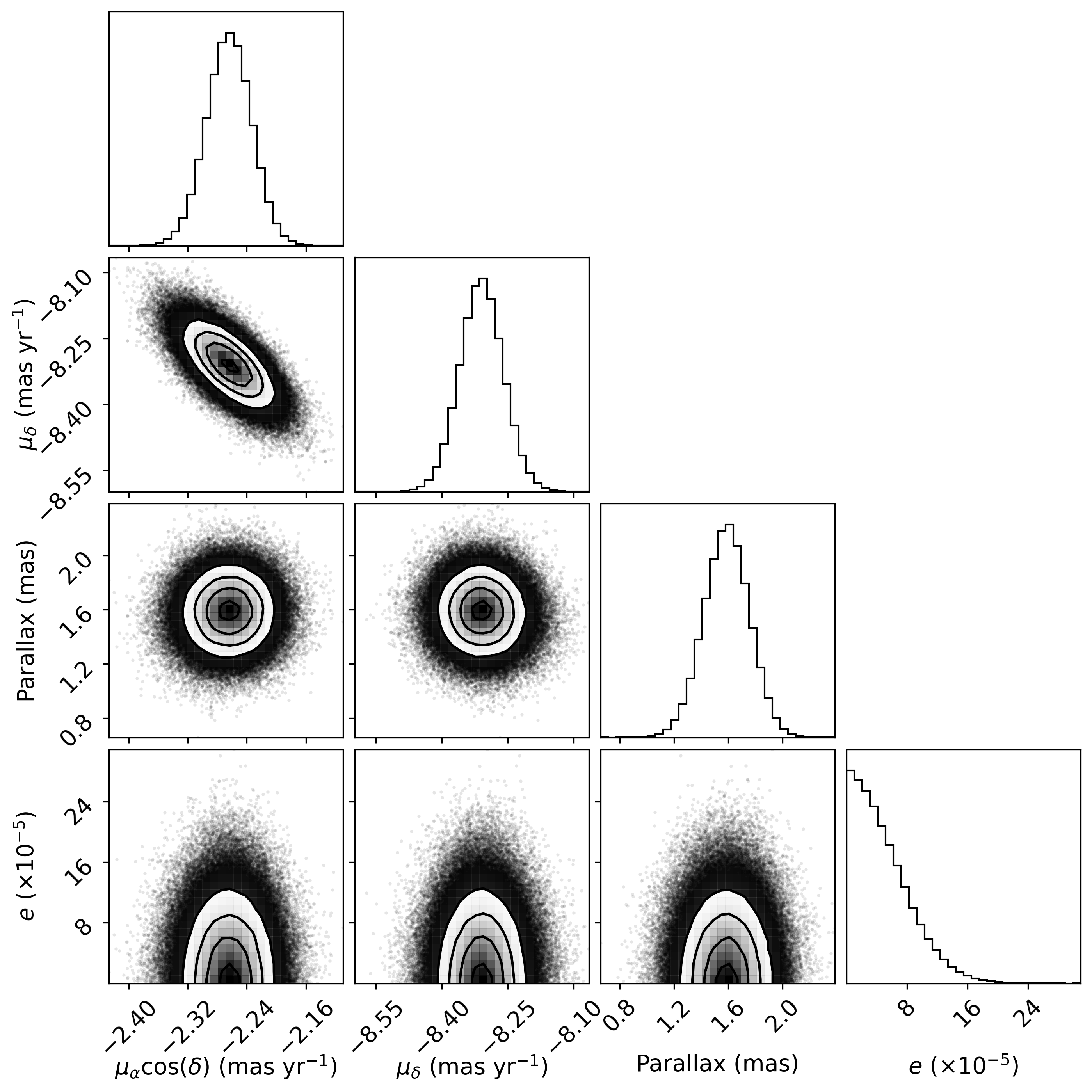}
    \caption{Posterior probability distributions for key pulsar parameters from timing of PSR~J2322$-$2650 with the MeerKAT radio telescope.}
    \label{fig:pulsar_corner_plot}
\end{figure}

\begin{table}[ht]
    \centering
    \caption{\textbf{Measured Timing Parameters of PSR~J2322$-$2650}}
    \begin{tabular}{ll}
    \hline
    \textbf{Parameter} & \textbf{Value} \\
    \hline
    \multicolumn{2}{c}{\textit{Pulsar Parameters}} \\
    RA, $\alpha$ (J2000) & 23:22:34.638364(4)\\
    DEC, $\delta$ (J2000)  & $-$26:50:58.39619(8)\\
    Proper motion in RA, $\mu_\alpha\cos\delta$  & $-$2.26(3)  mas yr$^{-1}$\\
    Proper motion in DEC, $\mu_\delta$ &  $-$8.31(5) mas yr$^{-1}$\\
    Spin Period ($P$) & 3.463099179266747(4) ms \\
    Period Derivative ($\dot{P}$) & $5.8586(16) \times 10^{-22}$ s s$^{-1}$ \\
    Epoch (MJD) & 59693.0\\
    Dispersion Measure & 6.14449(4) pc cm$^{-3}$\\
    Parallax & 1.59(17) mas \\
    Semi-major Axis$^*$ & 0.00278444 (s) \\
    Orbital Period & 0.3229640004(7) days \\
    Eccentricity, $e$ & $<$0.00012 (95\%)\\
    MJD Ascending Node & 59692.9699957(14)\\\multicolumn{2}{c}{\textit{Derived Parameters}} \\
    \hline
    Distance, $D$ & $630^{+75}_{-60}$ pc \\
    Transverse Velocity & $26 \pm 3$ km s$^{-1}$\\
    Intrinsic $\dot P$ & 1.8(4) $\times 10^{-22}$\\
    \hline
    \end{tabular}
     \begin{minipage}{0.5\linewidth}
     \centering
     $^*$Projected semi-major axis of the pulsar (not the companion), in light-seconds.
    \end{minipage}
    \label{tab:psrj2322}
\end{table}

We infer posterior probability distributions for the timing model parameters together with noise parameters using the Bayesian pulsar timing package inference \textsc{enterprise} \citep{enterprise}. We use a noise model that is standard in pulsar timing array analyses [including for the MPTA; \citealt{Miles+25}], which describes white noise, DM variations, and achromatic red noise in timing residuals. However, no significant red noise processes were detected. A corner plot \cite{corner} of the one- and two-dimensional marginal posterior probability distributions for the main timing model parameters of interest are shown in Fig.~\ref{fig:pulsar_corner_plot}. The measured timing model parameters are provided in Table \ref{tab:psrj2322}, including the pulsar parallax, proper motion, period, and period derivative. We do not detect the orbital eccentricity, and place a 95\% upper-limit of $e<0.00012$. The pulsar distance derived from the parallax is $D=630^{+75}_{-60}$\,pc and the transverse velocity with respect to the Sun is just 26\,km\,s$^{-1}$. Correcting for the peculiar motion of the Sun means that the velocity of the pulsar is small in the local standard of rest and indistinguishable from random motions in the Galaxy. The pulsar is experiencing the Shklovskii pseudo-acceleration due to its transverse velocity and distance, and this implies
that the intrinsic period derivative is $\dot P_{\rm i}=\dot P_{\rm obs}-PD\mu^2/c = 1.8(4) \times 10^{-22}$, where $\mu$ is the pulsar proper motion and $c$ is the speed of light. Such a low period derivative makes PSR~J2322$-$2650 one of the pulsars with the weakest magnetic field strength known ($B^2\propto P\dot P$), which may have helped the planet survive irradiation from the pulsar.

\section{Analysis of JWST data}
\label{sec:jwst_pipeline}
We extract one-dimensional spectra from the uncalibrated JWST data using a mixture of the JWST pipeline and our own software and methods.  The first stage of the pipeline consists in fitting a count rate to each pixel's measured sequence of accumulated counts.  While we follow the general approach of the JWST pipeline, we use our own software for every step of this stage.  Some of the changes in our approach have since been implemented in the official pipeline.

Correlated $1/f$ noise \citep{2010SPIE.7742E..1BM} is typically corrected by the JWST pipeline using non-light-sensitive reference pixels.  Unfortunately, our data are in subarray mode where most of these reference pixels are skipped rather than being read out.  Our treatment of $1/f$ noise differs somewhat from that of the JWST pipeline.  We describe that approach first, as we use it for all of our data processing including the construction of dark reference files.

\begin{figure}
    \centering\includegraphics[width=0.6\linewidth]{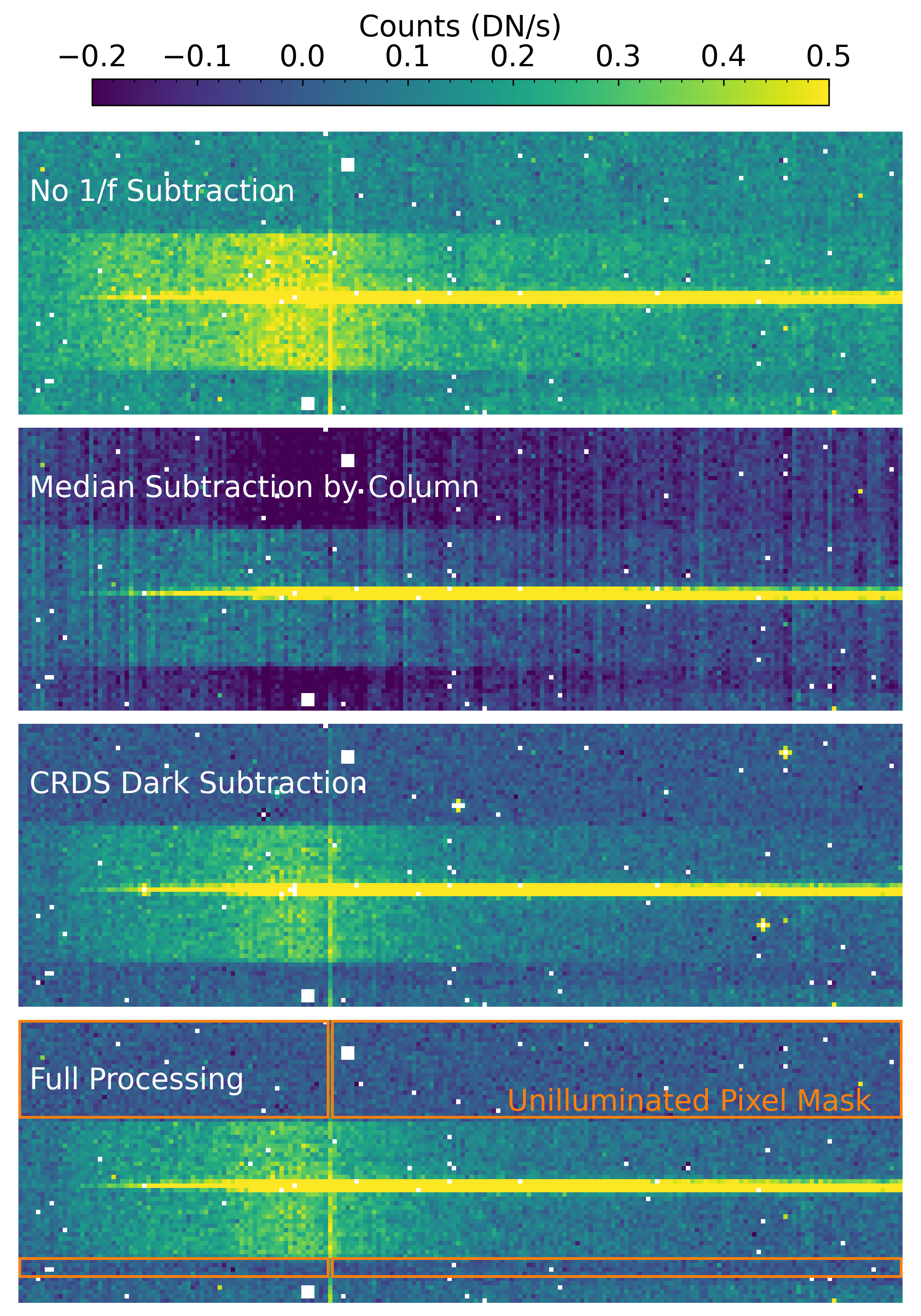}
    \caption{Comparison of different data reduction processes for the PRISM data, averaged over ten integrations. The region shown encompasses just under half of the full PRISM spectrum. 
    \label{fig:reduction_comparison}}
\end{figure}

The $1/f$ noise is correlated in time.  The pixels in the NIRSPEC subarrays are read out in columns, with each pixel in a column being read out sequentially before moving on to the next column.  As a result, temporally correlated noise appears as a spatial correlation, with the value of the noise being similar among all pixels in a column.  There are several approaches that can be used to mitigate this $1/f$ noise.  One is to compute and subtract the median value column-by-column from each read, while another is to determine a set of pixels that receive no signal (or nearly no signal) and to use these to construct a noise model.  Figure \ref{fig:reduction_comparison} shows both of these approaches, together with the result of doing no correction, in the top three panels.  In each case we show the count rate averaged over the same ten integrations with the same linear color scale.

The top panel of Figure \ref{fig:reduction_comparison} shows the result with no $1/f$ correction.  The background regions have a visibly nonzero count rate due to the low-frequency component of the $1/f$ that is shared by all reads; it would have a different value if a different set of integrations were averaged together.  The second panel shows the result of a median subtraction column-by-column.  It shows large single-column deviations from systematically different pixels along with biases from the fact that science data are not excluded.  

Our $1/f$ noise correction, shown in the third and fourth panels of Figure \ref{fig:reduction_comparison}, is based on the NSClean algorithm \citep{rauscher_2024} as implemented in the JWST pipeline \citep{bushouse_2025}.  Our changes to the algorithm consist of tuning the frequencies to be corrected and in designing a custom mask for the pixels to be used.  For PRISM data, we use four rows (numbers 7-10 of 64, beginning from 1) between shutters in addition to the top 22 rows.  These rows (with the exception of one bad column) are shown in the bottom panel of Figure \ref{fig:reduction_comparison}.  The frequencies to be corrected are specified by four numbers.  Frequencies below the lowest number or above the highest number (in Hz) are not corrected; frequencies between the second and third numbers are fully fit.  Frequencies between the first and second numbers, and between the third and fourth numbers, are apodized, i.e., fit with a weight that smoothly varies between zero and one.  The default values for these four frequencies are 1061, 1211, 49943, and 49957 Hz.  The latter number is just below 50,000 Hz; odd-even noise appears at a frequency of half the pixel rate of 100 kHz and is corrected.  We adopt values of 200, 600, 49943, and 49957 for our correction frequencies, i.e., we fit for lower frequency noise than the pipeline defaults.  We are able to do this because we have more pixels available for fitting in our custom mask than in the default mask.

In addition to the $1/f$ correction, we also revisit the JWST pipeline step of the dark subtraction.  The dark signal and bad pixels available at the JWST Calibration Reference Data System (CRDS\footnote{\url{https://jwst-crds.stsci.edu}}) have evolved since those data were taken, and do not provide a perfect match to our data.  Further, the dark signal for most pixels is extremely small, so that it is measured by the dark reference files only with low signal-to-noise ratios.  As a result, we do not perform a dark subtraction for most pixels. Any dark that we would construct or use from CRDS would have a realization of read noise in it, even if this read noise is averaged over many integrations.  Our mean spectra are, themselves, averaged over nearly 700 integrations, so the dark could easily make a significant contribution to the total read noise.  To avoid adding unnecessary read noise, we use darks only to correct the neighbors of hot pixels.

We define hot pixels for our purposes as those that accumulate more than 250 counts in 30 reads.  We do not aim to recover the value of the hot pixel but simply to remove the impact it makes on its neighbors due to interpixel capacitance.  To get this right pixel-by-pixel for our PRISM data, we construct a dark image for each read from 151 integrations of grating data with the same subarray configuration.  These data do have a faint visible trace.  We fit this trace with a quadratic in pixel coordinates.  Near the trace, the grating-derived dark would be an inappropriate choice as it is not truly dark.  We therefore substitute the value of the CRDS dark for pixels within 3 pixels of the center of the grating trace.

Finally, we use our hybrid dark to correct hot pixels and their immediate neighbors before masking the hot pixels themselves.  For an isolated hot pixel, we scale the dark to the science frame and subtract it in a $3\times3$ patch around the hot pixel.  The hot pixel itself now has zero flux; we mask it for all further analysis.  The immediate neighbors of the hot pixel now has had the effects of interpixel capacitance removed.  If a hot pixel is not isolated, we need to be careful not to doubly remove interpixel capacitance.  We use the same algorithm described above but, for pixels that border more than one hot pixel, we split their values across multiple $3\times3$ patches.  The bottom panel of Figure \ref{fig:reduction_comparison} shows the impact of the hot pixel processing, with neighbors of hot pixels appearing to be consistent with the large-scale background.

The steps above create a calibrated set of individual reads, but we still must fit for a count rate for each pixel in each integration.  For this we use the likelihood-based ramp fitting and jump detection algorithms described in \cite{brandt_2024a} and \cite{brandt_2024b}.  These were integrated into the JWST pipeline after these data were reduced.  So-called snowballs \citep{regan_2023} are clusters of pixels showing a jump in counts after an energetic cosmic ray hit.  Some of these jumps are at a level that is difficult to detect in an individual pixel.  To remove snowballs, we search for $7 \times 7$ pixel clusters where at least 80\% of the pixels show evidence of a jump.  We then compute the mean radius of this cluster of jumps, expand it by a factor of 2.5, and apply a mask to these reads for these pixels.  Finally, to mitigate persistence from strong cosmic ray hits, we exclude the remainder of the ramp for all pixels that had jumps registered on at least three consecutive reads.

Our suppression of correlated noise is good enough that the uncertainties in the ramps are now overestimated.  We quantify this by the distribution of the ratio of the ramp slope fit divided by its estimated uncertainty.  Because most pixels are either unilluminated or nearly so, this distribution should closely approximate a unit Gaussian.  Instead, we find that the distribution closely approximates a Gaussian with a standard deviation of 0.8.  We therefore reduce the estimated uncertainties on the ramp slopes by 20\%.

Our processing of the grating data matches our processing of the PRISM data as described above and shown in Figure \ref{fig:reduction_comparison}.  For the background pixels for a $1/f$ correction, we empirically determine the location of the trace by fitting a quadratic in pixel coordinates and exclude pixels within four pixels of the trace.  

The next step of our data reduction is spectral extraction.  We use the JWST pipeline for the initial step of flux calibration, continuing with our own routines subsequent to the generation of the {\tt calints} files.  We proceed with optimal extraction using an empirical trace profile.  For the PRISM data, the trace is nearly horizontal, lining up with one of the detector rows.  We limit the consideration of the trace to three pixels: the central pixel and its two nearest neighbors.  We first smooth the two-dimensional spectrum using a running median in the dispersion direction; we then normalize each column in the cross-dispersion direction.  We next fit a quadratic to the (normalized) pixel value as a function of position in the dispersion direction.  Finally, we renormalize this empirical trace profile at each position in the dispersion direction.

\begin{figure}
    \centering
    \includegraphics[width=0.6\linewidth]{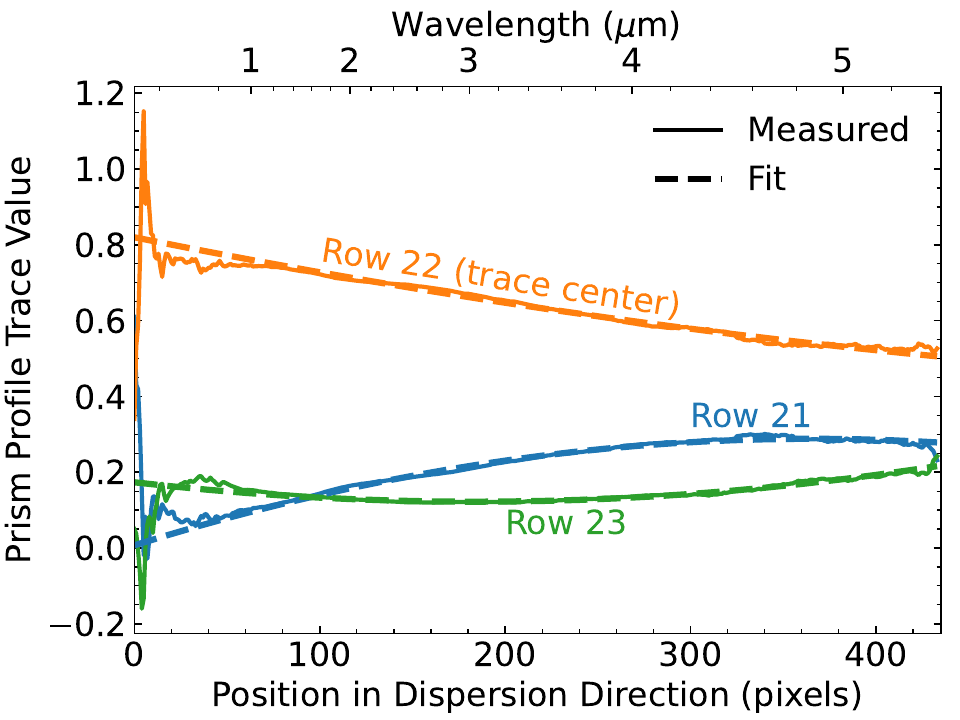}
    \caption{Empirical profile of the PRISM data used to extract spectra using the optimal extraction algorithm.  We use the three central pixels along the trace, which itself is nearly parallel to a detector row, to extract the spectrum.  The pixel values correspond to the JWST pipeline's {\tt x1dints} files; the corresponding wavelengths are given on the top axis.  Wavelengths below 0.8 \um are not used in the analysis.}
    \label{fig:prism_profile}
\end{figure}

Figure \ref{fig:prism_profile} shows the empirical trace profiles that we obtain from the JWST pipeline's {\tt calints} files using our custom {\tt rateints} files as input.  These consist of scaled two-dimensional, calibrated cutouts from the ramp slope files that we obtain as described earlier in this section.  We use the trace profiles in Figure \ref{fig:prism_profile} to perform an optimal extraction \cite{horne_1986}, but without weighting by the estimated uncertainties in each pixel.  The estimated uncertainties are correlated with the measurement errors themselves (fewer recorded photons leads to a lower estimated photon noise), so weighting the profile by the estimated uncertainties would introduce a bias.  The code that we use for spectral extraction, albeit not the code for constructing the trace profile, has also been added to the JWST pipeline since this reduction was done.  We use the JWST pipeline's wavelength calibration.

\begin{figure}
    \centering
    \includegraphics[width=0.6\linewidth]{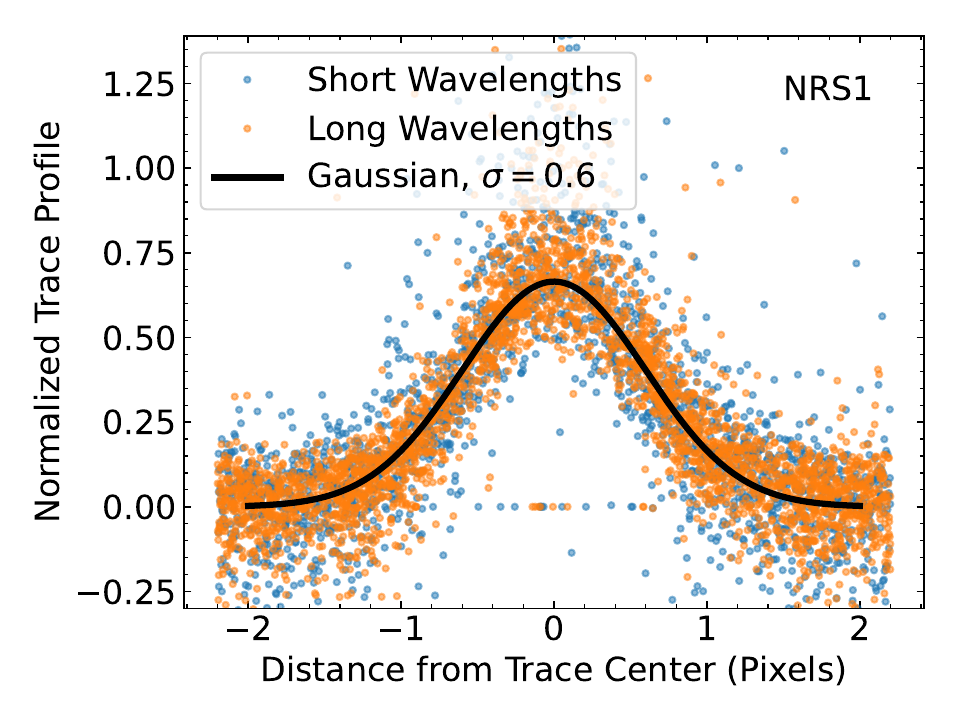}
    \includegraphics[width=0.6\linewidth]{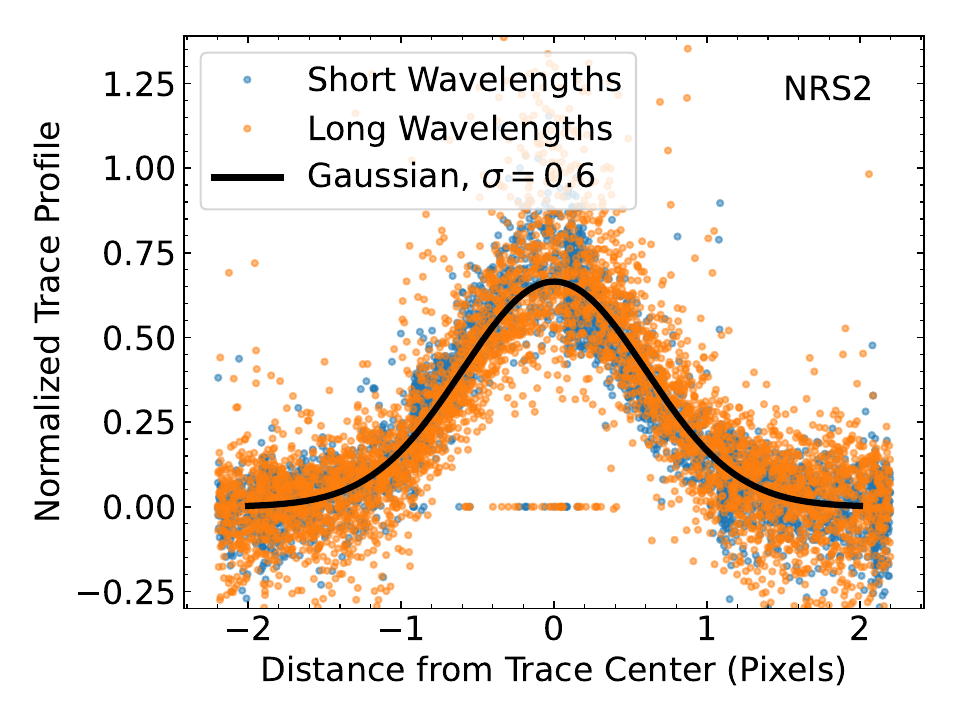}
    \caption{Adopted empirical profiles of the G235 grating traces for detector NRS1 (top) and NRS2 (bottom) shown with points for the measured pixel values after normalizing in the cross-dispersion direction and fitting the trace position.  The different colored points show the shorter- and longer-wavelength ranges of each detector.  }
    \label{fig:grating_traces}
\end{figure}

Our approach for grating spectral extraction differs for two reasons: the signal-to-noise is lower, and the trace shows significant curvature relative to the detector axes.  We adopt a multi-stage process.  We begin by averaging all of our grating exposures.  We then fit the trace profile using a quadratic in pixel coordinates.  Next, we normalize the average grating exposure column-by-column.  Finally, we plot each pixel value as a function of its distance from the trace.  A Gaussian with a standard deviation of 0.6 pixels provides a visually good fit.  Figure \ref{fig:grating_traces} shows this empirical trace profile together with the measured values as a function of distance from the trace center.  The trace profile appears to be consistent between the NRS1 and NRS2 detectors and does not appear to vary significantly with wavelength: the short-wavelength and long-wavelength halves of the data are consistent when plotted against one another.  A higher-fidelity reconstruction of the trace profile for the grating data could modestly improve our results.  Finally, we adopt the JWST pipeline's wavelength solution.

\section{Internal structure modeling}
\label{sec:interior_modelling}
We generate interior structure models \citep{nixon_2021,nixon_2024} of PSR\,J2322-2650\,b to explore its possible bulk composition. We consider three illustrative compositions for the planet: a pure carbon interior, a pure helium interior, and an interior of 99$\%$ helium and 1$\%$ carbon by mass.  We use temperature-dependent equations of state for helium \citep{chabrier_2021} and carbon \citep{kerley_2001}. For the mixed carbon and helium interior, we assume an additive volume law to obtain the mixed equation of state. We assume an isothermal-adiabatic temperature with a photospheric pressure of 1~mbar and radiative-convective boundary of 10~bar. The isothermal region of the atmosphere is fixed to 1500~K, with an adiabatic temperature profile used at deeper pressures.

Within the mass range of 1.5--2.5$M_{J}$, we find that a pure carbon object would have a radius of 0.393-0.394~$R_J$. Adjusting the temperature profile so that the radius becomes 1~$R_J$ requires a photospheric temperature of 5300~K, which yields a central temperature of 4.6$\times$10$^5$~K. This is significantly hotter than would be expected for PSR\,J2322-2650\,b. The pure helium and mixed helium + carbon models yield radii much closer to 1~$R_J$ using our nominal temperature profile. The pure helium models range from 0.837-0.947~$R_J$, and the mixed helium + carbon models range from 0.821-0.925~$R_J$, within the mass range of 1.5-2.5$M_{J}$.

Given the apparent depletion of hydrogen from the planet's atmosphere, and the fact that model planets with significant amounts of elements heavier than helium result in radii $\ll$1$R_J$, it seems likely from this initial analysis that the planet consists primarily of helium. A more complete analysis, using a wider range of compositions and accounting for the partly differentiated interior as well as the non-spherical shape, would be required to confirm this and refine constraints on the bulk composition. This detailed analysis is left for a future study.

\section{HELIOS models}
\label{sec:helios}
HELIOS is an open-source 1D radiative transfer code developed for planetary atmospheres \citep{malik_2017}.  Given planetary parameters and an atmospheric composition, it finds the radiative-convective equilibrium and outputs the resulting emission spectrum as well as the temperature-pressure profile.  HELIOS self-consistently supports external irradiation by main sequence stars, but we model the pulsar planet as an internally heated brown dwarf because most of the pulsar's radiation is in $\sim$GeV gamma rays, which penetrate well below the infrared photosphere.  In a helium atmosphere, the electromagnetic shower induced by the gamma ray collision deposits energy at a mass column density of $\Sigma \sim 2000 g$ cm$^{-2}$ (Equation 7 of \citealt{kandel_2020}), corresponding to a pressure of $P=\Sigma g = 10$ bar for the subpulsar point.  (This calculation neglects tidal forces from the pulsar, which reduce the effective g and therefore the characteristic P.)  We therefore turn off external irradiation in HELIOS by setting the stellar temperature to 0.

HELIOS requires opacity tables of all significant absorbers.  We generate correlated-k tables for C$_2$ and C$_2$H$_4$ using the ktable.py script included with HELIOS, starting from the high-resolution opacities in the DACE database \footnote{\url{https://dace.unige.ch/opacityDatabase/}}.  C$_2$H$_4$ is a stand-in for a generic molecule with carbon-hydrogen bonds.  C$_3$ is not in the DACE database, so we first generate opacities in DACE format from the line list \citep{lynas-gray_2024} using HELIOS-K \citep{grimm_2021}, before running ktable.py to produce the correlated-k table.

\begin{table}[htbp]
    \caption{\textbf{Gas volume mixing ratios, cloud parameters, and internal temperature adopted in HELIOS models}}
    \centering
    \setlength{\tabcolsep}{6pt}
    \begin{tabular}{c c c c}
    \hline
     & Day & Quadrature & Night \\
    \hline
    VMR$_{\rm He}$ & 0.994 & 0.997 & 0.9994\\  
    VMR$_{\rm C3}$ & 3\e{-3} & 1.5\e{-3} & 1\e{-4}\\  
    VMR$_{\rm C2}$ & 1\e{-3} & 5\e{-4} & 1\e{-4}\\  
    VMR$_{\rm C2H4}$ & 2\e{-3} & 1\e{-3} & 2\e{-4}\\  
    r$_{\rm dust}$ (\um) & 0.03 & 0.04 & 0.1\\
    T$_{\rm int}$ (K) & 2250 & 2000 & 1675 \\
    \hline
    \end{tabular}
    \label{table:helios_params}
\end{table}
 
For all our HELIOS models, we adopt a radius of 1 $R_J$ and a gravity of 4000 cm/s$^2$.  The atmosphere is assumed to be helium-dominated, with the gas abundances specified in Table \ref{table:helios_params}.  In our prescription, non-helium molecular abundances are lower on the night side than on the day side because larger molecules with unknown opacities (such as C$_4$ and C$_5$) are more abundant under chemical equilibrium, and because more carbon should condense into clouds.  All models have a graphite ``cloud'' (dust) extending upwards from 0.1 bar, with a cloud bottom mixing ratio of graphite particles of 1\e{-14} and a scale height 1.5$\times$ the gas scale height.  The only difference in cloud properties between day and night is in the particle radius, for which we prescribe 0.03 \um on the day side, 0.04 \um at quadrature, and 0.1 \um at night.  These radii were tuned to mute the spectral features, which we find to be much too strong in a clear atmosphere.  We also tune the one crucial parameter that controls the temperature structure of the atmosphere--$T_{\rm int}$, which quantifies the internal heat flux--so that the average brightness temperature of the resulting spectrum matches the observed brightness temperature.  A T/P profile more isothermal than predicted can also mute spectral features, reducing the need for a cloud/dust layer.  Given the large uncertainties in heating mechanism, abundances, and opacity sources, we do not consider HELIOS' T/P profile prediction to be robust.

These HELIOS models show that a cloudy helium-dominated atmosphere with trace amounts of C$_3$, C$_2$, and a hydrocarbon can reproduce the main features of the spectrum at all phases.  They are not intended to be a fully accurate model of the atmosphere.  Many parameters--such as the gas abundances, cloud particle mixing ratio, and cloud scale height--are simply assumed.  In addition, many potentially important physical processes are not modelled, such as the temperature inhomogeneity across the planetary disk; the non-spherical nature of the planet; and the non-constant effective gravity across the planet.  Finally, while we use C$_2$H$_4$ as a stand-in for a molecule with C-H bonds, we do not expect it to survive at 2000 K temperatures.  More likely candidates are C$_2$H and C$_3$H, which equilibrium chemistry calculations predict to be more abundant.  We will be able to include their opacities once high-temperature line lists are available.

Since HELIOS is a 1D plane-parallel model, it does not account for dayside inhomogeneity.  Irradiation, temperature, effective gravity, and viewing angle all vary significantly across the disk.  We will explore the impact of the plane-parallel assumption in future work.

\section{General circulation models}
\label{sec:gcms}
To model the atmosphere of the pulsar companion, we use the 3D General Circulation Model (GCM) \textbf{MITgcm} \citep{Adcroft2004} in the ADAM framework , which solves the primitive equations of meteorology on a cubed-sphere grid, and has previously been used to model hot gaseous planets extensively \citep{Showman2009, Lewis2014, Tan2019, Parmentier2021, Steinrueck2021, Roth2024, Tan2024}. These equations are coupled with a double-gray radiative transfer scheme (see \cite{Komacek2017} for details) routine with the thermal absorption coefficient calculated as a power-law in pressure \citep{Komacek2017}. Our irradiation scheme (described below) deposits a set total amount of heat in the atmosphere centered upon a prescribed pressure. As such, there is no shortwave heating in the double-gray radiative transfer scheme. 
We set the internal temperature, $T_{int}$ to be 500K, such that the net thermal flux emerging from the bottom of the domain is $\sigma T_{int}^{4}$.

To our knowledge, no GCM of a pulsar planet has been published previously (however, GCMs of neutron stars have been explored, see \citealt{Joonas2024}.)  As discussed in the previous subsection, the companion is primarily heated by the pulsar's GeV gamma rays, which deposit their energy at a typical pressure of $\sim$10 bar at the subpulsar point (and $\sim$10 cos($\theta$) bar at a angular distance $\theta$ from the subpulsar point).  We model this heating by artificially adding in a heat source at a characteristic pressure of $P_0\cos{\theta}$, where $P_0$=10 bar and $\theta$ is the angular distance from the substellar point.  The flux received from the pulsar is:

\begin{equation}
F = \frac{L_H}{4\pi a^2}\cos{\theta},
\end{equation}

where $L_H$ is the pulsar heating luminosity.  We adopt $L_H=9.1 \times 10^{32}$ erg/s, from the ICARUS light curve model fit.  We distribute this flux as a Gaussian in ln(P) space with a standard deviation of 1 (roughly corresponding to 1 scale height):

\begin{align}
    \frac{dF}{dlnP} = \frac{L_H}{4\pi a^2}\cos{\theta} \exp{\left(-\frac{1}{2}\ln^2\left[\frac{P}{P_0\cos{\theta}}\right]\right)} / \sqrt{2\pi}
\end{align}

To convert $\frac{dF}{dlnP}$ into a volumetric heating rate, we multiply by $|\frac{dlnP}{dz}|$.  In hydrostatic equilibrium, $P=P_a\exp{(-z/H)}$ over infinitesimally small dz, so $|\frac{dlnP}{dz}| = 1/H$.  We obtain a volumetric heating rate $\Gamma$ of:

\begin{align}
    \Gamma = \frac{L_H}{4\pi a^2}\cos{\theta} \frac{1}{\sqrt{2\pi}H}\exp{\left(-\frac{1}{2}\ln^2\left[\frac{P}{P_0\cos{\theta}}\right]\right)},
\end{align}

where H is the \textit{local} scale height.  Dividing $\Gamma$ by $\rho$ gives the heating rate per unit mass, which we put into the GCM following \cite{Showman2009} as a simple prescription for gamma ray heating.

As the physical properties of the companion are not well constrained, we take the mass and radius of the companion to be 1.5$M_{Jup}$ and 1 $R_{Jup}$ respectively. We use the published rotation period of 0.322 days \citep{spiewak_2017}. We use a dynamical timestep of 10 seconds for each simulation, which span from roughly 100 to $10^{-6}$ bars in pressure over 70 vertical levels.  We assume a mean molecular weight of 4 amu and set the heat capacity to that of a monoatomic gas, reflecting the likely helium-dominated composition of the atmosphere.  Each model was run for at least 1500 planetary days to ensure sufficient spin-up time.

\begin{figure}
    \centering
    \includegraphics[width=\linewidth]{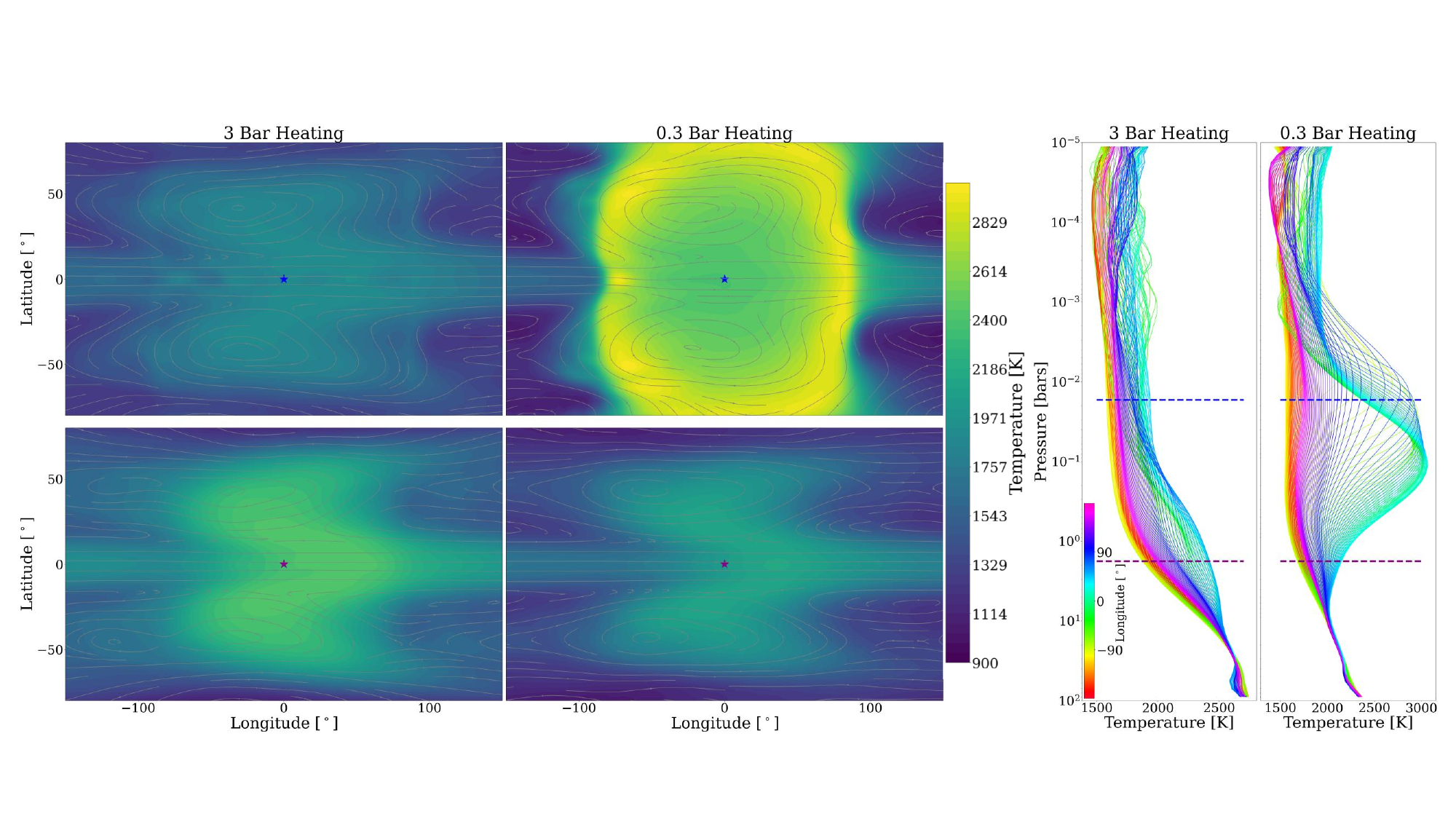}
    \caption{Temperature structure from 3D general circulation models. Here we show maps of time-averaged temperature and streamlines at two isobars (left) and equatorial temperature-pressure profiles (right) for two different heat deposition cases. The sub-pulsar point is indicated with a star. The isobaric projections correspond to roughly 20 mbar (top row of maps)---shown by the blue substellar marker and dashed line in the T-P profiles---and 2 bar (bottom row of maps)--corresponding to the purple dashed line and marker from our GCMs with a bespoke heating scheme given heat depositions centered at 3 bar  and 0.3 bar.  The combination of a rapid rotation rate and deep heating result in the circulation pattern. Specifically, we find westward shifted mid-latitude hot regions at the 2 bar pressure level for both cases of heat deposition. However, at 20 mbars, the model with more shallow heating shows a different temperature structure that is more symmetric and unlikely to produce a westward offset.  In the more deeply heated 3 bar case, the hot region being centered at negative longitudes implies a flux peak after phase 0.5, as observed.} 
    \label{fig:streamplot}
\end{figure}

We plot the isobaric temperature and streamlines for two pressure levels as well as equatorial temperature-pressure profiles in Figure \ref{fig:streamplot} for heating centered at 3 bar and 0.3 bar. The sub-pulsar point is indicated by the star, whose color shows the corresponding pressure level on the temperature-pressure profiles (purple for roughly 2 bar and blue for 20 mbar).  At 2 bar, both cases show mid-latitude hot anti-cyclones that are centered west of the substellar point. The resulting westward offset is a result of the rapid rotation of the planet; see \cite{Lee:2020aa,tan_2020,Zhan:2024aa} for previous studies of planets in this regime.  Because the pulsar planet is viewed at low inclination, the mid-latitudes contribute a disproportionate percentage of the total flux, further increasing the westward offset in the data. Notably, at 20 mbar, the more shallowly heated 0.3 bar model diverges from this circulation pattern, and appears unlikely to produce a westward offset. This suggests that this heating may be too shallow to explain the JWST phase curve of PSR J2322-2650b, and instead deep heating is required to match the observations. 

We note that our GCMs display stronger heat redistribution than in our observed phase curves (see Fig. \ref{fig:wlc}), particularly the 0.3 bar model.  Future work will improve the realism of the GCMs by exploring a larger model parameter space including further variation of the heat deposition, examining the influence of the infrared opacity, and self-consistently post-processing the model to produce spectroscopic phase curves to gauge the magnitude of the westward offset.

\section{Cross correlation}
\label{sec:cross_corr}

Cross correlation is used extensively to detect molecules in high resolution ($R \gtrsim 20,000$) transmission and emission spectra of exoplanets \citep{snellen_2025}.  JWST spectral resolutions are much lower, but cross correlation has nevertheless revealed CO in the NIRSpec G395H transmission spectrum of WASP-39b \citep{esparza-borges_2023}.

To check for the presence of a molecule in the atmosphere, we first make a template from -$\ln{(\sigma_\lambda)}$, where $\sigma_\lambda$ is the absorption cross section of the molecule at a fiducial temperature of 2000 K and 10 mbar (a rough approximation of the planet's infrared photospheric conditions).  The precise pressure is unimportant because thermal broadening dominates over pressure broadening at $P < 1$ bar, for the default pressure broadening assumed by ExoMol of $0.07\,\mathrm{cm}^{-1} (\frac{300 K}{2000 K})^{0.5} = 0.03\,\mathrm{cm}^{-1}$.  We generate the C$_2$ cross sections with ExoCross \citep{yurchenko_2018b} at R=$10^6$, take their natural log, broaden the result with a Gaussian filter to the maximum instrumental resolution (R=4200), interpolate onto a wavelength grid with a spacing of $\lambda/16,000$, and then detrend using a uniform filter with a width of 20 (equivalent to $\lambda/800)$.  This broadened and detrended -$\ln{(\sigma_\lambda)}$ is the cross correlation template.

We process the G235H data into a residuals grid using a methodology similar to that used in HRCCS.  We first compute the median spectrum across the whole observation and fit a second order polynomial to the region blueward of 3 \um (to avoid the C$_3$ cliff at 3.014 \um) to estimate the continuum.  We divide every spectrum by this polynomial, then divide it again by the median of the continuum-corrected spectrum to account for planetary flux variations over time.  Next, we subtract the column median from every column (corresponding to one wavelength).  This subtraction step removes all spectral or pseudo-spectral features due to bad pixels, the thermal background, or flatfielding errors, so long as the features do not vary with time.  The planet's spectral features are mostly protected from this subtraction because they shift over time due to orbital motion.  Finally, to match the processing done on the template, we detrend each row of the residuals grid with a uniform filter of width 20 and interpolate it onto the same wavelength grid.

\begin{figure*}[ht]
    \centering
    \begin{minipage}{0.49\textwidth}
        \centering
        \includegraphics[width=\textwidth]{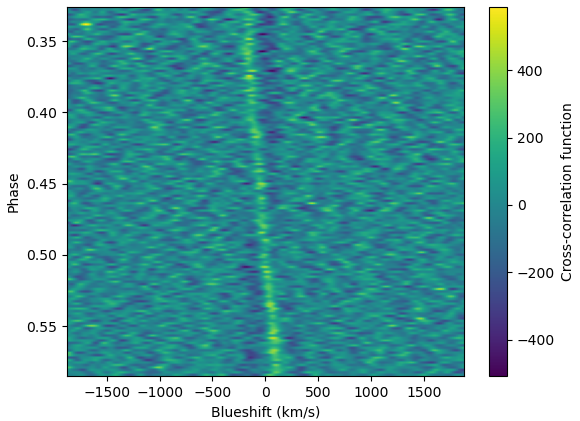}
    \end{minipage}%
    \hfill
    \begin{minipage}{0.49\textwidth}
        \centering
        \includegraphics[width=\textwidth]{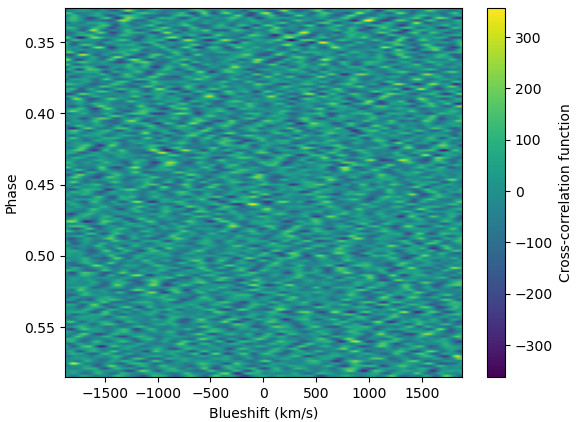}
    \end{minipage}
    \caption{Cross correlation value between the residuals grid and the C$_2$ template (left) or the CO template (right), as a function of time/phase and blueshift. The negative ``shadow'' on the left is from self-subtraction during the column-wise subtraction step.}
    \label{fig:ccf}
\end{figure*}

Figure \ref{fig:ccf} shows the cross correlation value as a function of time and blueshift.  C$_2$ is strongly detected, and the orbital motion of the planet is clearly visible.  Using exactly the same methodology, CO is not detected.  CO's absorption cross section is many orders of magnitude smaller than C$_2$'s throughout most of the G235H bandpass, but is only $\sim10\times$ smaller at 2.30--2.47 \um.  To improve sensitivity to CO, we tried only using this wavelength range, but still obtain no detection.  We do, however, detect C$_2$ at 6.1$\sigma$.


To derive a robust, model-independent upper limit on the CO-to-C$_2$ abundance ratio, we perform an analysis based solely on the cross sections of CO and C$_2$.  We model the brightness temperature as $T_b(\lambda) = T_0 - r\ln{\sigma(\lambda)}$, where $\sigma(\lambda) = f\sigma_{C2}(\lambda) + (1-f)\sigma_{CO}(\lambda)$ is the absorption cross section at 2000 K and 10 mbar.  $T_0$ is computed so that the median brightness temperature over 2.30--2.47 \um equals 2350 K (in accordance with observations).  We convert $T_b$ to $F_\nu$ with Planck's law, bin the $R=10^6$ $F_\nu$ down to an instrumental resolution of R=3100 with a Gaussian filter, interpolate to a wavelength grid with spacing of $\lambda/6000$ (equal to the G235H wavelength spacing at 2.39 \um), and add random Gaussian noise to each datapoint with a standard deviation equal to 8\% the mean $F_\nu$ (equal to the error on the real observations).  The cross correlation value at $V_{sys}=0$ of the detrended simulated observations with the detrended $\ln\sigma_{C2}(\lambda)$ ($\ln\sigma_{CO}(\lambda)$), divided by the standard deviation of the CCFs in the $1250 < |V_{sys}| < 5000$ km/s regions, is the detection significance of C$_2$ (CO).  For each f, we first tune $r$ so that the detection significance of C$_2$ reaches 6.0$\sigma$.  We then compute the detection significance of CO with the same $r$.  This exercise reveals that at f=1.0, $r=50 K$ is required to detect C$_2$ at 6.0$\sigma$, at which point CO is (unsurprisingly) ``detected'' at 0.1$\sigma$.  At f=0.25, $r$ must be 50 K, and CO is detected at 2.4$\sigma$.  The detection significance of CO increases to 3.0$\sigma$ at f=0.15 (r=75 K), and to 3.5$\sigma$ at f=0.1 (r=80 K).  These results are highly insensitive to both $r$ and the injected noise, provided that the combination of the two results in a 6$\sigma$ C$_2$ detection.  To prove this point, we double our injected noise from 8\% to 16\%, and find that at f=0.25, $r=140$ K achieves a 6.1$\sigma$ C$_2$ detection and a 2.4$\sigma$ CO detection--identical to our fiducial results.  We also halve our injected noise from 8\% to 4\% and find that at f=0.25, $r=35$ K results in an identical 6.1$\sigma$ C$_2$ detection and 2.4$\sigma$ CO detection.

\section{PRISM phase curve model}
We construct 3-D models of the heated companion surface in the Roche geometry, assuming that the pulsar heating is dominated by gamma-rays, as true for other black widows. Then, using the ICARUS binary light curve code \citep{breton_2012} with improvements including surface winds and more realistic gravity and limb darkening \citep{kandel_2020}, we integrate over the companion surface at each orbital phase, computing the emergent spectrum and light curve. Normally this employs a grid of stellar atmospheres, but at present we do not adequately understand the temperature/pressure variations of the very strong molecular carbon bands to compute reliable emergent spectra for the range of temperatures on the companion surface. Nevertheless the PRISM phase curves show clear thermal modulation, with near featureless Planckian spectra at minimum. Therefore we isolate (as most continuum dominated) two bands away from the strongest broad absorption features and fit these to light curve models, assuming Planck emission at the local effective temperature.

The light curve model fits for the companion's base (night time) temperature, the isotropic equivalent pulsar heating power, the system distance, and the orbital inclination.  We find that all fits converge to a high Roche lobe fill factor, and so fix the ratio of the nose radius to the L1 point distance to $f_1 =0.99$. The phase offset requires a significant surface wind, where the wind strength is characterized by the ratio of radiation to advection timescales $\epsilon=\tau_{\rm rad}/\tau_{\rm adv} \approx -0.12$; the negative sign indicates westward flow predominates at the latitudes which dominate the observed emission. While we use a standard analytic model for the limb darkening coefficients based on solar values \citep{neckel_2005} we find that the final results are sensitive to this assumption. Lacking detailed atmospheric emission models, we parameterize this sensitivity with a simple rescaling of the limb darkening amplitude.

This simple model does well at explaining the bulk of the light curve, indicates a low inclination, and gives a photometric distance consistent with the parallax constraint. The Earth-line-of-sight views shown below the curve show how even the night phase SED will be dominated by day-side fluxes from moderate latitude, viewed close to the limb. However, the model does not capture the nighttime variability of the planet from orbit to orbit--the night side is hotter the first time we observe it than the second, and by a larger amount at shorter wavelengths.  This may be from an eruptive event as seen in other black widow systems (e.g. \citealt{romani_2015}), although such eruptions are associated with very strong pulsar heating and/or magnetic activity on the companion, neither of which apply here. Alternatively, variable high-latitude winds at the terminator might cause fluctuations in emissivity or limb darkening. This variability at orbital minimum induces a large systematic which dominates our present inclination uncertainty, and hence pulsar mass uncertainty, exacerbated by our poor knowledge of the true limb darkening. We report parameters for fits to the full light curves, and for fits excluding the first or second minimum, in Table \ref{table:ICARUS_params}. Note that systematic errors dominate. The fit distance is at the large side of the parallax-allowed range; imposing this range as a prior makes little difference to best-fit parameters. The parameters are particularly sensitive to the limb darkening enhancement $C_{\rm LD}$; because the system is viewed at small inclination $i$, this allows a large range of inclinations, and hence of system masses. The PRISM data quality is such that, if reliable emergent spectra and limb darkening can be computed, we could fit to the full spectral range and substantially constrain the allowed parameters. Until better atmosphere models are available, SEDs covering several orbits would be required to understand the minimum variability and establish the true thermal minimum and precise orbital inclination.  

\begin{table}[htbp]
    \caption{Light Curve Fit Parameters. First errors are $\pm 1\sigma$ fit errors, inflated by $\chi^2$/DoF,\; second errors represent systematics from the range of the three fits.}
    \centering
    \setlength{\tabcolsep}{6pt}
    \begin{tabular}{l l l l}
    \hline
     & All Phases &-0.12-0.88 & 0.07-1.06\\
    \hline
    $T_N$ (K)                    & 904$\pm 15\pm40$   & 833   & 915 \\  
    $L_H$($10^{32}{\rm erg/s}$) & 9.1$\pm 0.2\pm2$   & 11.4   & 7.7 \\  
    $i$ ($^\circ$)               & 30.9$\pm 0.5\pm2.6$ & 28.1 & 33.1\\  
    $d$ (pc)                     & 750$\pm 10\pm45$   & 800   & 710\\  
    $\epsilon$                   & -0.123$\pm0.003$ &-0.133 & -0.124\\
     $C_{\rm LD}$                & 2.1$\pm0.2\pm0.8$  & 2.9 & 1.4 \\
%
     $\chi^2$/DOF                & 2.22  & 1.74 & 2.00\\
\hline
    $M_{\rm PSR}\,(M_\odot)$    &  1.97$\pm 0.1\pm0.5$ & 2.56  &  1.65\\
   
    \hline
    \end{tabular}
    \label{table:ICARUS_params}
\end{table}


At first sight, the large inferred isotropic heating power $L_H$ is puzzling. However, the pulsar gamma-ray emission, and to a lesser extent the particle flux, is concentrated toward the spin equator, especially for low magnetic inclination. Also, for large masses and the preferred stiff equation of state \citep{lattimer_2005} the pulsar moment of inertia is large, with $I_{45} \approx 1.2 (M/M_\odot)^{1.5}$, increasing the inferred pulsar spindown luminosity and magnetic field. For example with $M_{\rm PSR} = 2M_\odot$, we infer ${\dot E} \approx 6 \times 10^{32} {\rm erg~s}^{-1}$ and so the heating power should be beamed by $>1.5\times$ toward the equator. Thus the strong companion heating implies strong beaming, possibly in combination with a high pulsar mass.

Perhaps the most interesting system parameter is the pulsar mass $M_{\rm PSR}$, since a precise measurement should help us understand the accreted increment and the associated decrease in pulsar dipole field. If large, the mass might also contribute to our understanding of the dense matter equation of state. At present our fits are not constraining. However, as noted above, with improved modeling of the molecular carbon-dominated atmosphere, the JWST data can allow us to address these important questions.

\section{Equilibrium chemistry grid}
\label{sec:eq_chem_grid}
To derive conservative constraints on the elemental C/O and C/N ratios from our C$_2$/CO and C$_2$/CN lower limits, themselves derived from cross correlation of the G235H spectra, we generate two equilibrium chemistry abundance grids with FastChem.  The grids cover a log$_{10}$(C/He) of -5 to 1 in steps of 0.05, temperature of 600--4000 K in steps of 100 K, and pressure of $10^{-9}$ bar to $10^3$ bar in steps of 1 dex.  The first grid covers a log$_{10}$(C/O) of -1 to 5 in 60 steps, while the second grid covers a log$_{10}$(C/N) of 0 to 7 in 60 steps.  For reference, the solar photospheric values are log$_{10}$(C/O) = -0.22 and log$_{10}$(C/N) = 0.63.  The first grid only includes the elements He, C, and O, while the second grid only includes He, C, and N.

In the first grid, C$_2$/CO $>$ 0.17 is violated at all grid points unless C/O$>$2.  Restricting the temperature to $<$ 2300 K strengths the upper limit to C/O$>$11.  Further imposing the restrictions C/He $>$ 0.0035 (the solar value) and P $>$ 1 mbar raises the upper limit to C/O$>$100.

In the second grid, C$_2$/CN $>$ 32 is violated at all grid points unless C/N $>$ 300.  Restricting the temperature to $<$2300 K raises the limit to C/N $>$ 2000.  Further imposing the restrictions C/He $>$ 0.0082 (the solar value) and P $>$ 1 mbar raises the upper limit to C/N$>$10,000.

\bibliography{sample701}{}

@ARTICLE{koljonen_2025,
       author = {{Koljonen}, Karri I.~I. and {Linares}, Manuel},
        title = "{SpiderCat: A Catalog of Compact Binary Millisecond Pulsars}",
      journal = {arXiv e-prints},
         year = 2025,
        month = may,
          eid = {arXiv:2505.11691},
        pages = {arXiv:2505.11691},
          doi = {10.48550/arXiv.2505.11691},
archivePrefix = {arXiv},
       eprint = {2505.11691},
 primaryClass = {astro-ph.HE},
       adsurl = {https://ui.adsabs.harvard.edu/abs/2025arXiv250511691K}
}

@ARTICLE{Lee:2020aa,
       author = {{Lee}, Elspeth K.~H. and {Casewell}, Sarah L. and {Chubb}, Katy L. and {Hammond}, Mark and {Tan}, Xianyu and {Tsai}, Shang-Min and {Pierrehumbert}, Raymond T.},
        title = "{Simplified 3D GCM modelling of the irradiated brown dwarf WD 0137-349B}",
      journal = {\mnras},
         year = 2020,
        month = aug,
       volume = {496},
       number = {4},
        pages = {4674-4687},
          doi = {10.1093/mnras/staa1882},
archivePrefix = {arXiv},
       eprint = {2001.06558},
 primaryClass = {astro-ph.EP},
       adsurl = {https://ui.adsabs.harvard.edu/abs/2020MNRAS.496.4674L}
}

@article{Zhan:2024aa,
	adsurl = {https://ui.adsabs.harvard.edu/abs/2024ApJ...971..125Z},
	archiveprefix = {arXiv},
	author = {{Zhan}, Ruizhi and {Koll}, Daniel D.~B. and {Ding}, Feng},
	doi = {10.3847/1538-4357/ad54c1},
	eid = {125},
	eprint = {2406.03189},
	journal = {\apj},
	month = aug,
	number = {2},
	pages = {125},
	primaryclass = {astro-ph.EP},
	title = {{Novel Atmospheric Dynamics Shape the Inner Edge of the Habitable Zone around White Dwarfs}},
	volume = {971},
	year = 2024}

@ARTICLE{miles_2023,
       author = {{Miles}, M.~T. and {Shannon}, R.~M. and {Bailes}, M. and {Reardon}, D.~J. and {Keith}, M.~J. and {Cameron}, A.~D. and {Parthasarathy}, A. and {Shamohammadi}, M. and {Spiewak}, R. and {van Straten}, W. and {Buchner}, S. and {Camilo}, F. and {Geyer}, M. and {Karastergiou}, A. and {Kramer}, M. and {Serylak}, M. and {Theureau}, G. and {Venkatraman Krishnan}, V.},
        title = "{The MeerKAT Pulsar Timing Array: first data release}",
      journal = {\mnras},
         year = 2023,
        month = mar,
       volume = {519},
       number = {3},
        pages = {3976-3991},
          doi = {10.1093/mnras/stac3644},
archivePrefix = {arXiv},
       eprint = {2212.04648},
 primaryClass = {astro-ph.HE},
       adsurl = {https://ui.adsabs.harvard.edu/abs/2023MNRAS.519.3976M}
}

@ARTICLE{bailes_2020,
       author = {{Bailes}, M. and {Jameson}, A. and {Abbate}, F. and {Barr}, E.~D. and {Bhat}, N.~D.~R. and {Bondonneau}, L. and {Burgay}, M. and {Buchner}, S.~J. and {Camilo}, F. and {Champion}, D.~J. and {Cognard}, I. and {Demorest}, P.~B. and {Freire}, P.~C.~C. and {Gautam}, T. and {Geyer}, M. and {Griessmeier}, J. -M. and {Guillemot}, L. and {Hu}, H. and {Jankowski}, F. and {Johnston}, S. and {Karastergiou}, A. and {Karuppusamy}, R. and {Kaur}, D. and {Keith}, M.~J. and {Kramer}, M. and {van Leeuwen}, J. and {Lower}, M.~E. and {Maan}, Y. and {McLaughlin}, M.~A. and {Meyers}, B.~W. and {Os{\l}owski}, S. and {Oswald}, L.~S. and {Parthasarathy}, A. and {Pennucci}, T. and {Posselt}, B. and {Possenti}, A. and {Ransom}, S.~M. and {Reardon}, D.~J. and {Ridolfi}, A. and {Schollar}, C.~T.~G. and {Serylak}, M. and {Shaifullah}, G. and {Shamohammadi}, M. and {Shannon}, R.~M. and {Sobey}, C. and {Song}, X. and {Spiewak}, R. and {Stairs}, I.~H. and {Stappers}, B.~W. and {van Straten}, W. and {Szary}, A. and {Theureau}, G. and {Venkatraman Krishnan}, V. and {Weltevrede}, P. and {Wex}, N. and {Abbott}, T.~D. and {Adams}, G.~B. and {Burger}, J.~P. and {Gamatham}, R.~R.~G. and {Gouws}, M. and {Horn}, D.~M. and {Hugo}, B. and {Joubert}, A.~F. and {Manley}, J.~R. and {McAlpine}, K. and {Passmoor}, S.~S. and {Peens-Hough}, A. and {Ramudzuli}, Z.~R. and {Rust}, A. and {Salie}, S. and {Schwardt}, L.~C. and {Siebrits}, R. and {Van Tonder}, G. and {Van Tonder}, V. and {Welz}, M.~G.},
        title = "{The MeerKAT telescope as a pulsar facility: System verification and early science results from MeerTime}",
      journal = {Publications of the Astronomical Society of Australia},
         year = 2020,
        month = jul,
       volume = {37},
          eid = {e028},
        pages = {e028},
          doi = {10.1017/pasa.2020.19},
archivePrefix = {arXiv},
       eprint = {2005.14366},
 primaryClass = {astro-ph.IM},
       adsurl = {https://ui.adsabs.harvard.edu/abs/2020PASA...37...28B}
}

@article{spiewak_2017,
    author = {Spiewak, R and Bailes, M and Barr, E D and Bhat, N D R and Burgay, M and Cameron, A D and Champion, D J and Flynn, C M L and Jameson, A and Johnston, S and Keith, M J and Kramer, M and Kulkarni, S R and Levin, L and Lyne, A G and Morello, V and Ng, C and Possenti, A and Ravi, V and Stappers, B W and van Straten, W and Tiburzi, C},
    title = "{PSR~J2322-2650 - A low-luminosity millisecond pulsar with a planetary-mass companion}",
    journal = {Monthly Notices of the Royal Astronomical Society},
    volume = {475},
    number = {1},
    pages = {469-477},
    year = {2017},
    month = {12},
    issn = {0035-8711},
    doi = {10.1093/mnras/stx3157},
    url = {https://doi.org/10.1093/mnras/stx3157},
    eprint = {https://academic.oup.com/mnras/article-pdf/475/1/469/23534202/stx3157.pdf},
}

@ARTICLE{bell_2021,
       author = {{Bell}, Taylor J. and {Dang}, Lisa and {Cowan}, Nicolas B. and {Bean}, Jacob and {D{\'e}sert}, Jean-Michel and {Fortney}, Jonathan J. and {Keating}, Dylan and {Kempton}, Eliza and {Kreidberg}, Laura and {Line}, Michael R. and {Mansfield}, Megan and {Parmentier}, Vivien and {Stevenson}, Kevin B. and {Swain}, Mark and {Zellem}, Robert T.},
        title = "{A comprehensive reanalysis of Spitzer's 4.5 {\ensuremath{\mu}}m phase curves, and the phase variations of the ultra-hot Jupiters MASCARA-1b and KELT-16b}",
      journal = {\mnras},
         year = 2021,
        month = jul,
       volume = {504},
       number = {3},
        pages = {3316-3337},
          doi = {10.1093/mnras/stab1027},
archivePrefix = {arXiv},
       eprint = {2010.00687},
 primaryClass = {astro-ph.EP},
       adsurl = {https://ui.adsabs.harvard.edu/abs/2021MNRAS.504.3316B}
}

@ARTICLE{nixon_2021,
       author = {{Nixon}, Matthew C. and {Madhusudhan}, Nikku},
        title = "{How deep is the ocean? Exploring the phase structure of water-rich sub-Neptunes}",
      journal = {\mnras},
         year = 2021,
        month = aug,
       volume = {505},
       number = {3},
        pages = {3414-3432},
          doi = {10.1093/mnras/stab1500},
archivePrefix = {arXiv},
       eprint = {2106.02061},
 primaryClass = {astro-ph.EP},
       adsurl = {https://ui.adsabs.harvard.edu/abs/2021MNRAS.505.3414N}
}

@ARTICLE{chabrier_2021,
       author = {{Chabrier}, Gilles and {Debras}, Florian},
        title = "{A New Equation of State for Dense Hydrogen-Helium Mixtures. II. Taking into Account Hydrogen-Helium Interactions}",
      journal = {\apj},
         year = 2021,
        month = aug,
       volume = {917},
       number = {1},
          eid = {4},
        pages = {4},
          doi = {10.3847/1538-4357/abfc48},
archivePrefix = {arXiv},
       eprint = {2107.04434},
 primaryClass = {astro-ph.SR},
       adsurl = {https://ui.adsabs.harvard.edu/abs/2021ApJ...917....4C}
}

@ARTICLE{nixon_2024,
       author = {{Nixon}, Matthew C. and {Piette}, Anjali A.~A. and {Kempton}, Eliza M. -R. and {Gao}, Peter and {Bean}, Jacob L. and {Steinrueck}, Maria E. and {Mahajan}, Alexandra S. and {Eastman}, Jason D. and {Zhang}, Michael and {Rogers}, Leslie A.},
        title = "{New Insights into the Internal Structure of GJ 1214 b Informed by JWST}",
      journal = {\apjl},
         year = 2024,
        month = aug,
       volume = {970},
       number = {2},
          eid = {L28},
        pages = {L28},
          doi = {10.3847/2041-8213/ad615b},
archivePrefix = {arXiv},
       eprint = {2407.12079},
 primaryClass = {astro-ph.EP},
       adsurl = {https://ui.adsabs.harvard.edu/abs/2024ApJ...970L..28N}
}

@ARTICLE{dang_2018,
       author = {{Dang}, Lisa and {Cowan}, Nicolas B. and {Schwartz}, Joel C. and {Rauscher}, Emily and {Zhang}, Michael and {Knutson}, Heather A. and {Line}, Michael and {Dobbs-Dixon}, Ian and {Deming}, Drake and {Sundararajan}, Sudarsan and {Fortney}, Jonathan J. and {Zhao}, Ming},
        title = "{Detection of a westward hotspot offset in the atmosphere of hot gas giant CoRoT-2b}",
      journal = {Nature Astronomy},
         year = 2018,
        month = mar,
       volume = {2},
        pages = {220-227},
          doi = {10.1038/s41550-017-0351-6},
archivePrefix = {arXiv},
       eprint = {1801.06548},
 primaryClass = {astro-ph.EP},
       adsurl = {https://ui.adsabs.harvard.edu/abs/2018NatAs...2..220D}
}

@article{lynas-gray_2024,
    author = {Lynas-Gray, A E and Polyansky, O L and Tennyson, J and Yurchenko, S N and Zobov, N F},
    title = {ExoMol line lists – LXII. Ro-vibrational energy levels and line strengths for the propadienediylidene (C3) in its ground electronic state},
    journal = {Monthly Notices of the Royal Astronomical Society},
    volume = {535},
    number = {2},
    pages = {1439-1448},
    year = {2024},
    month = {10},
    issn = {0035-8711},
    doi = {10.1093/mnras/stae2425},
    url = {https://doi.org/10.1093/mnras/stae2425},
    eprint = {https://academic.oup.com/mnras/article-pdf/535/2/1439/60576055/stae2425.pdf},
}

@article{yurchenko_2018,
    author = {Yurchenko, Sergei N and Szabó, István and Pyatenko, Elizaveta and Tennyson, Jonathan},
    title = {ExoMol line lists XXXI: spectroscopy of lowest eights electronic states of C2},
    journal = {Monthly Notices of the Royal Astronomical Society},
    volume = {480},
    number = {3},
    pages = {3397-3411},
    year = {2018},
    month = {07},
    issn = {0035-8711},
    doi = {10.1093/mnras/sty2050},
    url = {https://doi.org/10.1093/mnras/sty2050},
    eprint = {https://academic.oup.com/mnras/article-pdf/480/3/3397/25508720/sty2050.pdf},
}

@ARTICLE{sing_2024,
       author = {{Sing}, David K. and {Evans-Soma}, Thomas M. and {Rustamkulov}, Zafar and {Lothringer}, Joshua D. and {Mayne}, Nathan J. and {Schlaufman}, Kevin C.},
        title = "{An Absolute Mass, Precise Age, and Hints of Planetary Winds for WASP-121A and b from a JWST NIRSpec Phase Curve}",
      journal = {\aj},
         year = 2024,
        month = dec,
       volume = {168},
       number = {6},
          eid = {231},
        pages = {231},
          doi = {10.3847/1538-3881/ad7fe7},
archivePrefix = {arXiv},
       eprint = {2501.03844},
 primaryClass = {astro-ph.EP},
       adsurl = {https://ui.adsabs.harvard.edu/abs/2024AJ....168..231S}
}

@report{kerley_2001,
    author = {Gerald I. Kerley and Lalit Chhabildas},
    title = {Multicomponent-Multiphase Equation of
State for Carbon},
    institution = {Sandia National Laboratories},
    year = {2001},
    url = {https://www.osti.gov/servlets/purl/787608}
}

@ARTICLE{taam_1986,
       author = {{Taam}, R.~E. and {van den Heuvel}, E.~P.~J.},
        title = "{Magnetic Field Decay and the Origin of Neutron Star Binaries}",
      journal = {\apj},
         year = 1986,
        month = jun,
       volume = {305},
        pages = {235},
          doi = {10.1086/164243},
       adsurl = {https://ui.adsabs.harvard.edu/abs/1986ApJ...305..235T}
}

@ARTICLE{saio_2002,
       author = {{Saio}, Hideyuki and {Jeffery}, C. Simon},
        title = "{Merged binary white dwarf evolution: rapidly accreting carbon-oxygen white dwarfs and the progeny of extreme helium stars}",
      journal = {\mnras},
         year = 2002,
        month = jun,
       volume = {333},
       number = {1},
        pages = {121-132},
          doi = {10.1046/j.1365-8711.2002.05384.x},
       adsurl = {https://ui.adsabs.harvard.edu/abs/2002MNRAS.333..121S}
}

@ARTICLE{webbink_1984,
       author = {{Webbink}, R.~F.},
        title = "{Double white dwarfs as progenitors of R Coronae Borealis stars and type I supernovae.}",
      journal = {\apj},
         year = 1984,
        month = feb,
       volume = {277},
        pages = {355-360},
          doi = {10.1086/161701},
       adsurl = {https://ui.adsabs.harvard.edu/abs/1984ApJ...277..355W}
}

@ARTICLE{abia_2008,
       author = {{Abia}, C. and {de Laverny}, P. and {Wahlin}, R.},
        title = "{Chemical analysis of carbon stars in the Local Group. II. The Carina dwarf spheroidal galaxy}",
      journal = {\aap},
         year = 2008,
        month = apr,
       volume = {481},
       number = {1},
        pages = {161-168},
          doi = {10.1051/0004-6361:20079114},
       adsurl = {https://ui.adsabs.harvard.edu/abs/2008A&A...481..161A}
}

@article{di_criscienzo_2013,
    author = {Di Criscienzo, M. and Dell’Agli, F. and Ventura, P. and Schneider, R. and Valiante, R. and La Franca, F. and Rossi, C. and Gallerani, S. and Maiolino, R.},
    title = {Dust formation in the winds of AGBs: the contribution at low metallicities},
    journal = {Monthly Notices of the Royal Astronomical Society},
    volume = {433},
    number = {1},
    pages = {313-323},
    year = {2013},
    month = {05},
    issn = {0035-8711},
    doi = {10.1093/mnras/stt732},
    url = {https://doi.org/10.1093/mnras/stt732},
    eprint = {https://academic.oup.com/mnras/article-pdf/433/1/313/18721049/stt732.pdf},
}

@ARTICLE{yurchenko_2018b,
       author = {{Yurchenko}, Sergei N. and {Al-Refaie}, Ahmed F. and {Tennyson}, Jonathan},
        title = "{EXOCROSS: a general program for generating spectra from molecular line lists}",
      journal = {\aap},
         year = 2018,
        month = jun,
       volume = {614},
          eid = {A131},
        pages = {A131},
          doi = {10.1051/0004-6361/201732531},
archivePrefix = {arXiv},
       eprint = {1801.09803},
 primaryClass = {astro-ph.EP},
       adsurl = {https://ui.adsabs.harvard.edu/abs/2018A&A...614A.131Y}
}

@ARTICLE{Lewis2014,
       author = {{Lewis}, Nikole K. and {Showman}, Adam P. and {Fortney}, Jonathan J. and {Knutson}, Heather A. and {Marley}, Mark S.},
        title = "{Atmospheric Circulation of Eccentric Hot Jupiter HAT-P-2b}",
      journal = {\apj},
         year = 2014,
        month = nov,
       volume = {795},
       number = {2},
          eid = {150},
        pages = {150},
          doi = {10.1088/0004-637X/795/2/150},
archivePrefix = {arXiv},
       eprint = {1409.5108},
 primaryClass = {astro-ph.EP},
       adsurl = {https://ui.adsabs.harvard.edu/abs/2014ApJ...795..150L}
}

@ARTICLE{Showman2009,
       author = {{Showman}, Adam P. and {Fortney}, Jonathan J. and {Lian}, Yuan and {Marley}, Mark S. and {Freedman}, Richard S. and {Knutson}, Heather A. and {Charbonneau}, David},
        title = "{Atmospheric Circulation of Hot Jupiters: Coupled Radiative-Dynamical General Circulation Model Simulations of HD 189733b and HD 209458b}",
      journal = {\apj},
         year = 2009,
        month = jul,
       volume = {699},
       number = {1},
        pages = {564-584},
          doi = {10.1088/0004-637X/699/1/564},
archivePrefix = {arXiv},
       eprint = {0809.2089},
 primaryClass = {astro-ph},
       adsurl = {https://ui.adsabs.harvard.edu/abs/2009ApJ...699..564S}
}

@ARTICLE{Tan2019,
       author = {{Tan}, Xianyu and {Komacek}, Thaddeus D.},
        title = "{The Atmospheric Circulation of Ultra-hot Jupiters}",
      journal = {\apj},
         year = 2019,
        month = nov,
       volume = {886},
       number = {1},
          eid = {26},
        pages = {26},
          doi = {10.3847/1538-4357/ab4a76},
archivePrefix = {arXiv},
       eprint = {1910.01622},
 primaryClass = {astro-ph.EP},
       adsurl = {https://ui.adsabs.harvard.edu/abs/2019ApJ...886...26T}
}

@ARTICLE{Parmentier2021,
       author = {{Parmentier}, Vivien and {Showman}, Adam P. and {Fortney}, Jonathan J.},
        title = "{The cloudy shape of hot Jupiter thermal phase curves}",
      journal = {\mnras},
         year = 2021,
        month = jan,
       volume = {501},
       number = {1},
        pages = {78-108},
          doi = {10.1093/mnras/staa3418},
archivePrefix = {arXiv},
       eprint = {2010.06934},
 primaryClass = {astro-ph.EP},
       adsurl = {https://ui.adsabs.harvard.edu/abs/2021MNRAS.501...78P}
}

@ARTICLE{Steinrueck2021,
       author = {{Steinrueck}, Maria E. and {Showman}, Adam P. and {Lavvas}, Panayotis and {Koskinen}, Tommi and {Tan}, Xianyu and {Zhang}, Xi},
        title = "{3D simulations of photochemical hazes in the atmosphere of hot Jupiter HD 189733b}",
      journal = {\mnras},
         year = 2021,
        month = jun,
       volume = {504},
       number = {2},
        pages = {2783-2799},
          doi = {10.1093/mnras/stab1053},
archivePrefix = {arXiv},
       eprint = {2011.14022},
 primaryClass = {astro-ph.EP},
       adsurl = {https://ui.adsabs.harvard.edu/abs/2021MNRAS.504.2783S}
}

@ARTICLE{Tan2024,
       author = {{Tan}, Xianyu and {Komacek}, Thaddeus D. and {Batalha}, Natasha E. and {Deming}, Drake and {Lupu}, Roxana and {Parmentier}, Vivien and {Pierrehumbert}, Raymond T.},
        title = "{Modelling the day-night temperature variations of ultra-hot Jupiters: confronting non-grey general circulation models and observations}",
      journal = {\mnras},
         year = 2024,
        month = feb,
       volume = {528},
       number = {1},
        pages = {1016-1036},
          doi = {10.1093/mnras/stae050},
archivePrefix = {arXiv},
       eprint = {2401.03859},
 primaryClass = {astro-ph.EP},
       adsurl = {https://ui.adsabs.harvard.edu/abs/2024MNRAS.528.1016T}
}

@ARTICLE{Roth2024,
       author = {{Roth}, Alexander and {Parmentier}, Vivien and {Hammond}, Mark},
        title = "{Hot Jupiter diversity and the onset of TiO/VO revealed by a large grid of non-grey global circulation models}",
      journal = {\mnras},
         year = 2024,
        month = jun,
       volume = {531},
       number = {1},
        pages = {1056-1083},
          doi = {10.1093/mnras/stae984},
archivePrefix = {arXiv},
       eprint = {2404.09626},
 primaryClass = {astro-ph.EP},
       adsurl = {https://ui.adsabs.harvard.edu/abs/2024MNRAS.531.1056R}
}

@ARTICLE{Joonas2024,
       author = {{N{\"a}ttil{\"a}}, Joonas and {Cho}, James Y. -K. and {Skinner}, Jack W. and {Most}, Elias R. and {Ripperda}, Bart},
        title = "{Neutron Star Atmosphere{\textendash}Ocean Dynamics}",
      journal = {\apj},
         year = 2024,
        month = aug,
       volume = {971},
       number = {1},
          eid = {37},
        pages = {37},
          doi = {10.3847/1538-4357/ad54c2},
archivePrefix = {arXiv},
       eprint = {2306.08186},
 primaryClass = {astro-ph.HE},
       adsurl = {https://ui.adsabs.harvard.edu/abs/2024ApJ...971...37N}
}

@ARTICLE{malik_2017,
       author = {{Malik}, Matej and {Grosheintz}, Luc and {Mendon{\c{c}}a}, Jo{\~a}o M. and {Grimm}, Simon L. and {Lavie}, Baptiste and {Kitzmann}, Daniel and {Tsai}, Shang-Min and {Burrows}, Adam and {Kreidberg}, Laura and {Bedell}, Megan and {Bean}, Jacob L. and {Stevenson}, Kevin B. and {Heng}, Kevin},
        title = "{HELIOS: An Open-source, GPU-accelerated Radiative Transfer Code for Self-consistent Exoplanetary Atmospheres}",
      journal = {\aj},
         year = 2017,
        month = feb,
       volume = {153},
       number = {2},
          eid = {56},
        pages = {56},
          doi = {10.3847/1538-3881/153/2/56},
archivePrefix = {arXiv},
       eprint = {1606.05474},
 primaryClass = {astro-ph.EP},
       adsurl = {https://ui.adsabs.harvard.edu/abs/2017AJ....153...56M}
}

@ARTICLE{malik_2019,
       author = {{Malik}, Matej and {Kitzmann}, Daniel and {Mendon{\c{c}}a}, Jo{\~a}o M. and {Grimm}, Simon L. and {Marleau}, Gabriel-Dominique and {Linder}, Esther F. and {Tsai}, Shang-Min and {Heng}, Kevin},
        title = "{Self-luminous and Irradiated Exoplanetary Atmospheres Explored with HELIOS}",
      journal = {\aj},
         year = 2019,
        month = may,
       volume = {157},
       number = {5},
          eid = {170},
        pages = {170},
          doi = {10.3847/1538-3881/ab1084},
archivePrefix = {arXiv},
       eprint = {1903.06794},
 primaryClass = {astro-ph.EP},
       adsurl = {https://ui.adsabs.harvard.edu/abs/2019AJ....157..170M}
}

@ARTICLE{Adcroft2004,
       author = {{Adcroft}, Alistair and {Campin}, Jean-Michel and {Hill}, Chris and {Marshall}, John},
        title = "{Implementation of an Atmosphere Ocean General Circulation Model on the Expanded Spherical Cube}",
      journal = {Monthly Weather Review},
         year = 2004,
        month = jan,
       volume = {132},
       number = {12},
        pages = {2845},
          doi = {10.1175/MWR2823.1},
       adsurl = {https://ui.adsabs.harvard.edu/abs/2004MWRv..132.2845A}
}

@ARTICLE{Komacek2017,
       author = {{Komacek}, Thaddeus D. and {Showman}, Adam P. and {Tan}, Xianyu},
        title = "{Atmospheric Circulation of Hot Jupiters: Dayside-Nightside Temperature Differences. II. Comparison with Observations}",
      journal = {\apj},
         year = 2017,
        month = feb,
       volume = {835},
       number = {2},
          eid = {198},
        pages = {198},
          doi = {10.3847/1538-4357/835/2/198},
archivePrefix = {arXiv},
       eprint = {1610.03893},
 primaryClass = {astro-ph.EP},
       adsurl = {https://ui.adsabs.harvard.edu/abs/2017ApJ...835..198K}
}

@ARTICLE{romani_2016,
       author = {{Romani}, Roger W. and {Graham}, Melissa L. and {Filippenko}, Alexei V. and {Zheng}, WeiKang},
        title = "{PSR J1301+0833: A Kinematic Study of a Black-widow Pulsar}",
      journal = {\apj},
         year = 2016,
        month = dec,
       volume = {833},
       number = {2},
          eid = {138},
        pages = {138},
          doi = {10.3847/1538-4357/833/2/138},
       adsurl = {https://ui.adsabs.harvard.edu/abs/2016ApJ...833..138R}
}

@ARTICLE{romani_2015,
       author = {{Romani}, Roger W. and {Filippenko}, Alexei V. and {Cenko}, S. Bradley},
        title = "{A Spectroscopic Study of the Extreme Black Widow PSR J1311-3430}",
      journal = {\apj},
         year = 2015,
        month = may,
       volume = {804},
       number = {2},
          eid = {115},
        pages = {115},
          doi = {10.1088/0004-637X/804/2/115},
archivePrefix = {arXiv},
       eprint = {1503.05247},
 primaryClass = {astro-ph.HE},
       adsurl = {https://ui.adsabs.harvard.edu/abs/2015ApJ...804..115R}
}

@ARTICLE{1990Natur.347..741R,
       author = {{Romani}, Roger W.},
        title = "{A unified model of neutron-star magnetic fields}",
      journal = {\nat},
         year = 1990,
        month = oct,
       volume = {347},
       number = {6295},
        pages = {741-743},
          doi = {10.1038/347741a0},
       adsurl = {https://ui.adsabs.harvard.edu/abs/1990Natur.347..741R}
}

@ARTICLE{1974SvA....18..217B,
       author = {{Bisnovatyi-Kogan}, G.~S. and {Komberg}, B.~V.},
        title = "{Pulsars and close binary systems}",
      journal = {Soviet Astronomy},
         year = 1974,
        month = oct,
       volume = {18},
        pages = {217},
       adsurl = {https://ui.adsabs.harvard.edu/abs/1974SvA....18..217B}
}

@ARTICLE{2017JApA...38...48M,
       author = {{Mukherjee}, Dipanjan},
        title = "{Revisiting Field Burial by Accretion onto Neutron Stars}",
      journal = {Journal of Astrophysics and Astronomy},
         year = 2017,
        month = sep,
       volume = {38},
       number = {3},
          eid = {48},
        pages = {48},
          doi = {10.1007/s12036-017-9465-6},
archivePrefix = {arXiv},
       eprint = {1709.07332},
 primaryClass = {astro-ph.HE},
       adsurl = {https://ui.adsabs.harvard.edu/abs/2017JApA...38...48M}
}

@misc{enterprise,
  author       = {Justin A. Ellis and Michele Vallisneri and Stephen R. Taylor and Paul T. Baker},
  title        = {ENTERPRISE: Enhanced Numerical Toolbox Enabling a Robust PulsaR Inference SuitE},
  month        = sep,
  year         = 2020,
  howpublished = {Zenodo},
  doi          = {10.5281/zenodo.4059815},
  url          = {https://doi.org/10.5281/zenodo.4059815}
}

@article{corner,
  doi = {10.21105/joss.00024},
  url = {https://doi.org/10.21105/joss.00024},
  year  = {2016},
  month = {jun},
  publisher = {The Open Journal},
  volume = {1},
  number = {2},
  pages = {24},
  author = {Daniel Foreman-Mackey},
  title = {corner.py: Scatterplot matrices in Python},
  journal = {The Journal of Open Source Software}
}

@ARTICLE{Miles+25,
       author = {{Miles}, Matthew T. and {Shannon}, Ryan M. and {Reardon}, Daniel J. and {Bailes}, Matthew and {Champion}, David J. and {Geyer}, Marisa and {Gitika}, Pratyasha and {Grunthal}, Kathrin and {Keith}, Michael J. and {Kramer}, Michael and {Kulkarni}, Atharva D. and {Nathan}, Rowina S. and {Parthasarathy}, Aditya and {Porayko}, Nataliya K. and {Singha}, Jaikhomba and {Theureau}, Gilles and {Abbate}, Federico and {Buchner}, Sarah and {Cameron}, Andrew D. and {Camilo}, Fernando and {Moreschi}, Beatrice E. and {Shaifullah}, Golam and {Shamohammadi}, Mohsen and {Krishnan}, Vivek Venkatraman},
        title = "{The MeerKAT Pulsar Timing Array: the 4.5-yr data release and the noise and stochastic signals of the millisecond pulsar population}",
      journal = {\mnras},
         year = 2025,
        month = jan,
       volume = {536},
       number = {2},
        pages = {1467-1488},
          doi = {10.1093/mnras/stae2572},
archivePrefix = {arXiv},
       eprint = {2412.01148},
 primaryClass = {astro-ph.HE},
       adsurl = {https://ui.adsabs.harvard.edu/abs/2025MNRAS.536.1467M}
}

@ARTICLE{Hobbs+06,
       author = {{Hobbs}, G.~B. and {Edwards}, R.~T. and {Manchester}, R.~N.},
        title = "{TEMPO2, a new pulsar-timing package - I. An overview}",
      journal = {\mnras},
         year = 2006,
        month = jun,
       volume = {369},
       number = {2},
        pages = {655-672},
          doi = {10.1111/j.1365-2966.2006.10302.x},
archivePrefix = {arXiv},
       eprint = {astro-ph/0603381},
 primaryClass = {astro-ph},
       adsurl = {https://ui.adsabs.harvard.edu/abs/2006MNRAS.369..655H}
}

@ARTICLE{bailes_2011,
       author = {{Bailes}, M. and {Bates}, S.~D. and {Bhalerao}, V. and {Bhat}, N.~D.~R. and {Burgay}, M. and {Burke-Spolaor}, S. and {D'Amico}, N. and {Johnston}, S. and {Keith}, M.~J. and {Kramer}, M. and {Kulkarni}, S.~R. and {Levin}, L. and {Lyne}, A.~G. and {Milia}, S. and {Possenti}, A. and {Spitler}, L. and {Stappers}, B. and {van Straten}, W.},
        title = "{Transformation of a Star into a Planet in a Millisecond Pulsar Binary}",
      journal = {Science},
         year = 2011,
        month = sep,
       volume = {333},
       number = {6050},
        pages = {1717},
          doi = {10.1126/science.1208890},
archivePrefix = {arXiv},
       eprint = {1108.5201},
 primaryClass = {astro-ph.SR},
       adsurl = {https://ui.adsabs.harvard.edu/abs/2011Sci...333.1717B}
}

@ARTICLE{kandel_2020,
       author = {{Kandel}, D. and {Romani}, Roger W.},
        title = "{Atmospheric Circulation on Black Widow Companions}",
      journal = {\apj},
         year = 2020,
        month = apr,
       volume = {892},
       number = {2},
          eid = {101},
        pages = {101},
          doi = {10.3847/1538-4357/ab7b62},
archivePrefix = {arXiv},
       eprint = {2002.12483},
 primaryClass = {astro-ph.HE},
       adsurl = {https://ui.adsabs.harvard.edu/abs/2020ApJ...892..101K}
}

@ARTICLE{breton_2012,
       author = {{Breton}, R.~P. and {Rappaport}, S.~A. and {van Kerkwijk}, M.~H. and {Carter}, J.~A.},
        title = "{KOI 1224: A Fourth Bloated Hot White Dwarf Companion Found with Kepler}",
      journal = {\apj},
         year = 2012,
        month = apr,
       volume = {748},
       number = {2},
          eid = {115},
        pages = {115},
          doi = {10.1088/0004-637X/748/2/115},
archivePrefix = {arXiv},
       eprint = {1109.6847},
 primaryClass = {astro-ph.SR},
       adsurl = {https://ui.adsabs.harvard.edu/abs/2012ApJ...748..115B}
}

@ARTICLE{snellen_2025,
       author = {{Snellen}, Ignas},
        title = "{Exoplanet atmospheres at high spectral resolution}",
      journal = {arXiv e-prints},
         year = 2025,
        month = may,
          eid = {arXiv:2505.08926},
        pages = {arXiv:2505.08926},
          doi = {10.48550/arXiv.2505.08926},
archivePrefix = {arXiv},
       eprint = {2505.08926},
 primaryClass = {astro-ph.EP},
       adsurl = {https://ui.adsabs.harvard.edu/abs/2025arXiv250508926S}
}

@ARTICLE{tan_2020,
       author = {{Tan}, Xianyu and {Showman}, Adam P.},
        title = "{Atmospheric Circulation of Tidally Locked Gas Giants with Increasing Rotation and Implications for White Dwarf-Brown Dwarf Systems}",
      journal = {\apj},
         year = 2020,
        month = oct,
       volume = {902},
       number = {1},
          eid = {27},
        pages = {27},
          doi = {10.3847/1538-4357/abb3d4},
archivePrefix = {arXiv},
       eprint = {2001.06269},
 primaryClass = {astro-ph.EP},
       adsurl = {https://ui.adsabs.harvard.edu/abs/2020ApJ...902...27T}
}

@ARTICLE{grimm_2021,
       author = {{Grimm}, Simon L. and {Malik}, Matej and {Kitzmann}, Daniel and {Guzm{\'a}n-Mesa}, Andrea and {Hoeijmakers}, H. Jens and {Fisher}, Chloe and {Mendon{\c{c}}a}, Jo{\~a}o M. and {Yurchenko}, Sergey N. and {Tennyson}, Jonathan and {Alesina}, Fabien and {Buchschacher}, Nicolas and {Burnier}, Julien and {Segransan}, Damien and {Kurucz}, Robert L. and {Heng}, Kevin},
        title = "{HELIOS-K 2.0 Opacity Calculator and Open-source Opacity Database for Exoplanetary Atmospheres}",
      journal = {\apjs},
         year = 2021,
        month = mar,
       volume = {253},
       number = {1},
          eid = {30},
        pages = {30},
          doi = {10.3847/1538-4365/abd773},
archivePrefix = {arXiv},
       eprint = {2101.02005},
 primaryClass = {astro-ph.EP},
       adsurl = {https://ui.adsabs.harvard.edu/abs/2021ApJS..253...30G}
}

@ARTICLE{rauscher_2024,
       author = {{Rauscher}, Bernard J.},
        title = "{NSClean: An Algorithm for Removing Correlated Noise from JWST NIRSpec Images}",
      journal = {\pasp},
         year = 2024,
        month = jan,
       volume = {136},
       number = {1},
          eid = {015001},
        pages = {015001},
          doi = {10.1088/1538-3873/ad1b36},
archivePrefix = {arXiv},
       eprint = {2306.03250},
 primaryClass = {astro-ph.IM},
       adsurl = {https://ui.adsabs.harvard.edu/abs/2024PASP..136a5001R}
}

@software{bushouse_2025,
       author = {{Bushouse}, Howard and {Eisenhamer}, Jonathan and {Dencheva}, Nadia and {Davies}, James and {Greenfield}, Perry and {Morrison}, Jane and {Hodge}, Phil and {Simon}, Bernie and {Grumm}, David and {Droettboom}, Michael and {Slavich}, Edward and {Sosey}, Megan and {Pauly}, Tyler and {Miller}, Todd and {Jedrzejewski}, Robert and {Hack}, Warren and {Davis}, David and {Crawford}, Steven and {Law}, David and {Gordon}, Karl and {Regan}, Michael and {Cara}, Mihai and {MacDonald}, Ken and {Bradley}, Larry and {Shanahan}, Clare and {Jamieson}, William and {Teodoro}, Mairan and {Williams}, Thomas and {Pena-Guerrero}, Maria and {Graham}, Brett and {Molter}, Edward and {Brandt}, Timothy and {Hayes}, Christian and {Cooper}, Rachel and {Clarke}, Melanie and {Filippazzo}, Joseph},
        title = "{JWST Calibration Pipeline}",
         year = 2025,
        month = apr,
          eid = {10.5281/zenodo.6984365},
          doi = {10.5281/zenodo.6984365},
      version = {1.18.0},
    publisher = {Zenodo},
       adsurl = {https://ui.adsabs.harvard.edu/abs/2023zndo...6984365B}
}

@ARTICLE{horne_1986,
       author = {{Horne}, K.},
        title = "{An optimal extraction algorithm for CCD spectroscopy.}",
      journal = {\pasp},
         year = 1986,
        month = jun,
       volume = {98},
        pages = {609-617},
          doi = {10.1086/131801},
       adsurl = {https://ui.adsabs.harvard.edu/abs/1986PASP...98..609H}
}

@ARTICLE{brandt_2024a,
       author = {{Brandt}, Timothy D.},
        title = "{Optimal Fitting and Debiasing for Detectors Read Out Up-the-Ramp}",
      journal = {\pasp},
         year = 2024,
        month = apr,
       volume = {136},
       number = {4},
          eid = {045004},
        pages = {045004},
          doi = {10.1088/1538-3873/ad38d9},
archivePrefix = {arXiv},
       eprint = {2309.08753},
 primaryClass = {astro-ph.IM},
       adsurl = {https://ui.adsabs.harvard.edu/abs/2024PASP..136d5004B}
}

@ARTICLE{brandt_2024b,
       author = {{Brandt}, Timothy D.},
        title = "{Likelihood-based Jump Detection and Cosmic Ray Rejection for Detectors Read Out Up-the-ramp}",
      journal = {\pasp},
         year = 2024,
        month = apr,
       volume = {136},
       number = {4},
          eid = {045005},
        pages = {045005},
          doi = {10.1088/1538-3873/ad38da},
archivePrefix = {arXiv},
       eprint = {2404.01326},
 primaryClass = {astro-ph.IM},
       adsurl = {https://ui.adsabs.harvard.edu/abs/2024PASP..136d5005B}
}

@MISC{regan_2023,
       author = {{Regan}, Michael},
        title = "{Detection and Flagging of Showers and Snowballs in JWST}",
 howpublished = {Technical Report JWST-STScI-008545},
         year = 2023,
        month = sep,
        pages = {8545},
       adsurl = {https://ui.adsabs.harvard.edu/abs/2023jwst.rept.8545R}
}

@ARTICLE{neckel_2005,
       author = {{Neckel}, Heinz},
        title = "{Analytical Reference Functions F({\ensuremath{\lambda}}) for the Sun's Limb Darkening and Its Absolute Continuum Intensities ({\ensuremath{\lambda}}{\ensuremath{\lambda}} 300 to 1100 m)}",
      journal = {Solar Physics},
         year = 2005,
        month = jun,
       volume = {229},
       number = {1},
        pages = {13-33},
          doi = {10.1007/s11207-005-4081-z},
       adsurl = {https://ui.adsabs.harvard.edu/abs/2005SoPh..229...13N}
}

@INPROCEEDINGS{2010SPIE.7742E..1BM,
       author = {{Moseley}, S.~H. and {Arendt}, Richard G. and {Fixsen}, D.~J. and {Lindler}, Don and {Loose}, Markus and {Rauscher}, Bernard J.},
        title = "{Reducing the read noise of H2RG detector arrays: eliminating correlated noise with efficient use of reference signals}",
    booktitle = {High Energy, Optical, and Infrared Detectors for Astronomy IV},
         year = 2010,
       editor = {{Holland}, Andrew D. and {Dorn}, David A.},
       series = {Society of Photo-Optical Instrumentation Engineers (SPIE) Conference Series},
       volume = {7742},
        month = jul,
          eid = {77421B},
        pages = {77421B},
          doi = {10.1117/12.866773},
       adsurl = {https://ui.adsabs.harvard.edu/abs/2010SPIE.7742E..1BM}
}

@ARTICLE{lattimer_2005,
       author = {{Lattimer}, James M. and {Schutz}, Bernard F.},
        title = "{Constraining the Equation of State with Moment of Inertia Measurements}",
      journal = {\apj},
         year = 2005,
        month = aug,
       volume = {629},
       number = {2},
        pages = {979-984},
          doi = {10.1086/431543},
archivePrefix = {arXiv},
       eprint = {astro-ph/0411470},
 primaryClass = {astro-ph},
       adsurl = {https://ui.adsabs.harvard.edu/abs/2005ApJ...629..979L}
}

@ARTICLE{shklovskii_1970,
       author = {{Shklovskii}, I.~S.},
        title = "{Possible Causes of the Secular Increase in Pulsar Periods.}",
      journal = {Soviet Astronomy},
         year = 1970,
        month = feb,
       volume = {13},
        pages = {562},
       adsurl = {https://ui.adsabs.harvard.edu/abs/1970SvA....13..562S}
}

@ARTICLE{shamohammadi_2024,
       author = {{Shamohammadi}, M. and {Bailes}, M. and {Flynn}, C. and {Reardon}, D.~J. and {Shannon}, R.~M. and {Buchner}, S. and {Cameron}, A.~D. and {Camilo}, F. and {Corongiu}, A. and {Geyer}, M. and {Kramer}, M. and {Miles}, M. and {Spiewak}, R.},
        title = "{MeerKAT Pulsar Timing Array parallaxes and proper motions}",
      journal = {\mnras},
         year = 2024,
        month = may,
       volume = {530},
       number = {1},
        pages = {287-306},
          doi = {10.1093/mnras/stae016},
archivePrefix = {arXiv},
       eprint = {2401.06963},
 primaryClass = {astro-ph.HE},
       adsurl = {https://ui.adsabs.harvard.edu/abs/2024MNRAS.530..287S}
}

@article{tammer_2004,
  author  = {Tammer, M.},
  title   = {G. Sokrates: Infrared and Raman characteristic group frequencies: tables and charts},
  journal = {Colloid and Polymer Science},
  year    = {2004},
  month   = dec,
  volume  = {283},
  number  = {2},
  pages   = {235},
  issn    = {1435-1536},
  doi     = {10.1007/s00396-004-1164-6},
  url     = {https://doi.org/10.1007/s00396-004-1164-6}
}

@PHDTHESIS{hansen_1996,
       author = {{Hansen}, Bradley Miles Stougaard},
        title = "{The Ages, Speeds and Offspring of Pulsars}",
       school = {California Institute of Technology},
         year = 1996,
        month = jan,
       adsurl = {https://ui.adsabs.harvard.edu/abs/1996PhDT.........5H}
}

@ARTICLE{esparza-borges_2023,
       author = {{Esparza-Borges}, Emma and {L{\'o}pez-Morales}, Mercedes and {Adams Redai}, J{\'e}a I. and {Pall{\'e}}, Enric and {Kirk}, James and {Casasayas-Barris}, N{\'u}ria and {Batalha}, Natasha E. and {Rackham}, Benjamin V. and {Bean}, Jacob L. and {Casewell}, S.~L. and {Decin}, Leen and {Dos Santos}, Leonardo A. and {Garc{\'\i}a Mu{\~n}oz}, Antonio and {Harrington}, Joseph and {Heng}, Kevin and {Hu}, Renyu and {Mancini}, Luigi and {Molaverdikhani}, Karan and {Morello}, Giuseppe and {Nikolov}, Nikolay K. and {Nixon}, Matthew C. and {Redfield}, Seth and {Stevenson}, Kevin B. and {Wakeford}, Hannah R. and {Alam}, Munazza K. and {Benneke}, Bj{\"o}rn and {Blecic}, Jasmina and {Crouzet}, Nicolas and {Daylan}, Tansu and {Inglis}, Julie and {Kreidberg}, Laura and {Petit dit de la Roche}, Dominique J.~M. and {Turner}, Jake D.},
        title = "{Detection of Carbon Monoxide in the Atmosphere of WASP-39b Applying Standard Cross-correlation Techniques to JWST NIRSpec G395H Data}",
      journal = {\apjl},
         year = 2023,
        month = sep,
       volume = {955},
       number = {1},
          eid = {L19},
        pages = {L19},
          doi = {10.3847/2041-8213/acf27b},
archivePrefix = {arXiv},
       eprint = {2309.00036},
 primaryClass = {astro-ph.EP},
       adsurl = {https://ui.adsabs.harvard.edu/abs/2023ApJ...955L..19E}
}

@ARTICLE{tennyson_2018,
       author = {{Tennyson}, Jonathan and {Yurchenko}, Sergei N.},
        title = "{The ExoMol Atlas of Molecular Opacities}",
      journal = {Atoms},
         year = 2018,
        month = may,
       volume = {6},
       number = {2},
          eid = {26},
        pages = {26},
          doi = {10.3390/atoms6020026},
archivePrefix = {arXiv},
       eprint = {1805.03711},
 primaryClass = {astro-ph.SR},
       adsurl = {https://ui.adsabs.harvard.edu/abs/2018Atoms...6...26T}
}

@ARTICLE{lattimer_2012,
       author = {{Lattimer}, James M.},
        title = "{The Nuclear Equation of State and Neutron Star Masses}",
      journal = {Annual Review of Nuclear and Particle Science},
         year = 2012,
        month = nov,
       volume = {62},
       number = {1},
        pages = {485-515},
          doi = {10.1146/annurev-nucl-102711-095018},
archivePrefix = {arXiv},
       eprint = {1305.3510},
 primaryClass = {nucl-th},
       adsurl = {https://ui.adsabs.harvard.edu/abs/2012ARNPS..62..485L}
}

@ARTICLE{ballet_2023,
       author = {{Ballet}, J. and {Bruel}, P. and {Burnett}, T.~H. and {Lott}, B. and {The Fermi-LAT collaboration}},
        title = "{Fermi Large Area Telescope Fourth Source Catalog Data Release 4 (4FGL-DR4)}",
      journal = {arXiv e-prints},
         year = 2023,
        month = jul,
          eid = {arXiv:2307.12546},
        pages = {arXiv:2307.12546},
          doi = {10.48550/arXiv.2307.12546},
archivePrefix = {arXiv},
       eprint = {2307.12546},
 primaryClass = {astro-ph.HE},
       adsurl = {https://ui.adsabs.harvard.edu/abs/2023arXiv230712546B}
}

@article{draghis_2019,
    doi = {10.3847/1538-4357/ab378b},
    url = {https://dx.doi.org/10.3847/1538-4357/ab378b},
    year = {2019},
    month = {sep},
    publisher = {The American Astronomical Society},
    volume = {883},
    number = {1},
    pages = {108},
    author = {Draghis, Paul and Romani, Roger W. and Filippenko, Alexei V. and Brink, Thomas G. and Zheng, WeiKang and Halpern, Jules P. and Camilo, Fernando},
    title = {Multiband Optical Light Curves of Black-widow Pulsars},
    journal = {The Astrophysical Journal}
}

@ARTICLE{philippov_2020,
       author = {{Philippov}, Alexander and {Timokhin}, Andrey and {Spitkovsky}, Anatoly},
        title = "{Origin of Pulsar Radio Emission}",
      journal = {Physical Review Letters},
         year = 2020,
        month = jun,
       volume = {124},
       number = {24},
          eid = {245101},
        pages = {245101},
          doi = {10.1103/PhysRevLett.124.245101},
archivePrefix = {arXiv},
       eprint = {2001.02236},
 primaryClass = {astro-ph.HE},
       adsurl = {https://ui.adsabs.harvard.edu/abs/2020PhRvL.124x5101P}
}

@ARTICLE{van_straten_2012,
       author = {{van Straten}, Willem and {Demorest}, Paul and {Oslowski}, Stefan},
        title = "{Pulsar Data Analysis with PSRCHIVE}",
      journal = {Astronomical Research and Technology},
         year = 2012,
        month = jul,
       volume = {9},
       number = {3},
        pages = {237-256},
          doi = {10.48550/arXiv.1205.6276},
archivePrefix = {arXiv},
       eprint = {1205.6276},
 primaryClass = {astro-ph.IM},
       adsurl = {https://ui.adsabs.harvard.edu/abs/2012AR&T....9..237V}
}

@ARTICLE{madhusudhan_2012,
       author = {{Madhusudhan}, Nikku},
        title = "{C/O Ratio as a Dimension for Characterizing Exoplanetary Atmospheres}",
      journal = {\apj},
         year = 2012,
        month = oct,
       volume = {758},
       number = {1},
          eid = {36},
        pages = {36},
          doi = {10.1088/0004-637X/758/1/36},
archivePrefix = {arXiv},
       eprint = {1209.2412},
 primaryClass = {astro-ph.EP},
       adsurl = {https://ui.adsabs.harvard.edu/abs/2012ApJ...758...36M}
}

@ARTICLE{stock_2018,
       author = {{Stock}, Joachim W. and {Kitzmann}, Daniel and {Patzer}, A. Beate C. and {Sedlmayr}, Erwin},
        title = "{FastChem: A computer program for efficient complex chemical equilibrium calculations in the neutral/ionized gas phase with applications to stellar and planetary atmospheres}",
      journal = {\mnras},
         year = 2018,
        month = sep,
       volume = {479},
       number = {1},
        pages = {865-874},
          doi = {10.1093/mnras/sty1531},
archivePrefix = {arXiv},
       eprint = {1804.05010},
 primaryClass = {astro-ph.EP},
       adsurl = {https://ui.adsabs.harvard.edu/abs/2018MNRAS.479..865S}
}

@ARTICLE{stock_2022,
       author = {{Stock}, Joachim W. and {Kitzmann}, Daniel and {Patzer}, A. Beate C.},
        title = "{FASTCHEM 2 : an improved computer program to determine the gas-phase chemical equilibrium composition for arbitrary element distributions}",
      journal = {\mnras},
         year = 2022,
        month = dec,
       volume = {517},
       number = {3},
        pages = {4070-4080},
          doi = {10.1093/mnras/stac2623},
archivePrefix = {arXiv},
       eprint = {2206.08247},
 primaryClass = {astro-ph.EP},
       adsurl = {https://ui.adsabs.harvard.edu/abs/2022MNRAS.517.4070S}
}

@ARTICLE{mehla_2025,
       author = {{Mehla}, Advait and {Kasliwal}, Mansi M. and {Karambelkar}, Viraj and {Tisserand}, Patrick and {Crawford}, Courtney and {Clayton}, Geoffrey and {Soon}, Jamie and {Bhalerao}, Varun},
        title = "{Oxygen Isotope Ratios in Hydrogen-deficient Carbon Stars: A Correlation with Effective Temperature and Implications for White Dwarf Merger Outcomes}",
      journal = {\pasp},
         year = 2025,
        month = apr,
       volume = {137},
       number = {4},
          eid = {044201},
        pages = {044201},
          doi = {10.1088/1538-3873/adc0bf},
archivePrefix = {arXiv},
       eprint = {2412.03664},
 primaryClass = {astro-ph.SR},
       adsurl = {https://ui.adsabs.harvard.edu/abs/2025PASP..137d4201M}
}

@ARTICLE{benvenuto_2012,
       author = {{Benvenuto}, O.~G. and {De Vito}, M.~A. and {Horvath}, J.~E.},
        title = "{Evolutionary Trajectories of Ultracompact ``Black Widow'' Pulsars with Very Low Mass Companions}",
      journal = {\apjl},
         year = 2012,
        month = jul,
       volume = {753},
       number = {2},
          eid = {L33},
        pages = {L33},
          doi = {10.1088/2041-8205/753/2/L33},
archivePrefix = {arXiv},
       eprint = {1206.2389},
 primaryClass = {astro-ph.SR},
       adsurl = {https://ui.adsabs.harvard.edu/abs/2012ApJ...753L..33B}
}

@article{astropy:2013,
Adsurl = {http://adsabs.harvard.edu/abs/2013A%26A...558A..33A},
Archiveprefix = {arXiv},
Author = {{Astropy Collaboration} and {Robitaille}, T.~P. and {Tollerud}, E.~J. and {Greenfield}, P. and {Droettboom}, M. and {Bray}, E. and {Aldcroft}, T. and {Davis}, M. and {Ginsburg}, A. and {Price-Whelan}, A.~M. and {Kerzendorf}, W.~E. and {Conley}, A. and {Crighton}, N. and {Barbary}, K. and {Muna}, D. and {Ferguson}, H. and {Grollier}, F. and {Parikh}, M.~M. and {Nair}, P.~H. and {Unther}, H.~M. and {Deil}, C. and {Woillez}, J. and {Conseil}, S. and {Kramer}, R. and {Turner}, J.~E.~H. and {Singer}, L. and {Fox}, R. and {Weaver}, B.~A. and {Zabalza}, V. and {Edwards}, Z.~I. and {Azalee Bostroem}, K. and {Burke}, D.~J. and {Casey}, A.~R. and {Crawford}, S.~M. and {Dencheva}, N. and {Ely}, J. and {Jenness}, T. and {Labrie}, K. and {Lim}, P.~L. and {Pierfederici}, F. and {Pontzen}, A. and {Ptak}, A. and {Refsdal}, B. and {Servillat}, M. and {Streicher}, O.},
Doi = {10.1051/0004-6361/201322068},
Eid = {A33},
Eprint = {1307.6212},
Journal = {\aap},
Month = oct,
Pages = {A33},
Primaryclass = {astro-ph.IM},
Title = {{Astropy: A community Python package for astronomy}},
Volume = 558,
Year = 2013}

@ARTICLE{astropy:2018,
       author = {{Astropy Collaboration} and {Price-Whelan}, A.~M. and
         {Sip{\H{o}}cz}, B.~M. and {G{\"u}nther}, H.~M. and {Lim}, P.~L. and
         {Crawford}, S.~M. and {Conseil}, S. and {Shupe}, D.~L. and
         {Craig}, M.~W. and {Dencheva}, N. and {Ginsburg}, A. and {Vand
        erPlas}, J.~T. and {Bradley}, L.~D. and {P{\'e}rez-Su{\'a}rez}, D. and
         {de Val-Borro}, M. and {Aldcroft}, T.~L. and {Cruz}, K.~L. and
         {Robitaille}, T.~P. and {Tollerud}, E.~J. and {Ardelean}, C. and
         {Babej}, T. and {Bach}, Y.~P. and {Bachetti}, M. and {Bakanov}, A.~V. and
         {Bamford}, S.~P. and {Barentsen}, G. and {Barmby}, P. and
         {Baumbach}, A. and {Berry}, K.~L. and {Biscani}, F. and {Boquien}, M. and
         {Bostroem}, K.~A. and {Bouma}, L.~G. and {Brammer}, G.~B. and
         {Bray}, E.~M. and {Breytenbach}, H. and {Buddelmeijer}, H. and
         {Burke}, D.~J. and {Calderone}, G. and {Cano Rodr{\'\i}guez}, J.~L. and
         {Cara}, M. and {Cardoso}, J.~V.~M. and {Cheedella}, S. and {Copin}, Y. and
         {Corrales}, L. and {Crichton}, D. and {D'Avella}, D. and {Deil}, C. and
         {Depagne}, {\'E}. and {Dietrich}, J.~P. and {Donath}, A. and
         {Droettboom}, M. and {Earl}, N. and {Erben}, T. and {Fabbro}, S. and
         {Ferreira}, L.~A. and {Finethy}, T. and {Fox}, R.~T. and
         {Garrison}, L.~H. and {Gibbons}, S.~L.~J. and {Goldstein}, D.~A. and
         {Gommers}, R. and {Greco}, J.~P. and {Greenfield}, P. and
         {Groener}, A.~M. and {Grollier}, F. and {Hagen}, A. and {Hirst}, P. and
         {Homeier}, D. and {Horton}, A.~J. and {Hosseinzadeh}, G. and {Hu}, L. and
         {Hunkeler}, J.~S. and {Ivezi{\'c}}, {\v{Z}}. and {Jain}, A. and
         {Jenness}, T. and {Kanarek}, G. and {Kendrew}, S. and {Kern}, N.~S. and
         {Kerzendorf}, W.~E. and {Khvalko}, A. and {King}, J. and {Kirkby}, D. and
         {Kulkarni}, A.~M. and {Kumar}, A. and {Lee}, A. and {Lenz}, D. and
         {Littlefair}, S.~P. and {Ma}, Z. and {Macleod}, D.~M. and
         {Mastropietro}, M. and {McCully}, C. and {Montagnac}, S. and
         {Morris}, B.~M. and {Mueller}, M. and {Mumford}, S.~J. and {Muna}, D. and
         {Murphy}, N.~A. and {Nelson}, S. and {Nguyen}, G.~H. and
         {Ninan}, J.~P. and {N{\"o}the}, M. and {Ogaz}, S. and {Oh}, S. and
         {Parejko}, J.~K. and {Parley}, N. and {Pascual}, S. and {Patil}, R. and
         {Patil}, A.~A. and {Plunkett}, A.~L. and {Prochaska}, J.~X. and
         {Rastogi}, T. and {Reddy Janga}, V. and {Sabater}, J. and
         {Sakurikar}, P. and {Seifert}, M. and {Sherbert}, L.~E. and
         {Sherwood-Taylor}, H. and {Shih}, A.~Y. and {Sick}, J. and
         {Silbiger}, M.~T. and {Singanamalla}, S. and {Singer}, L.~P. and
         {Sladen}, P.~H. and {Sooley}, K.~A. and {Sornarajah}, S. and
         {Streicher}, O. and {Teuben}, P. and {Thomas}, S.~W. and
         {Tremblay}, G.~R. and {Turner}, J.~E.~H. and {Terr{\'o}n}, V. and
         {van Kerkwijk}, M.~H. and {de la Vega}, A. and {Watkins}, L.~L. and
         {Weaver}, B.~A. and {Whitmore}, J.~B. and {Woillez}, J. and
         {Zabalza}, V. and {Astropy Contributors}},
        title = "{The Astropy Project: Building an Open-science Project and Status of the v2.0 Core Package}",
      journal = {\aj},
         year = 2018,
        month = sep,
       volume = {156},
       number = {3},
          eid = {123},
        pages = {123},
          doi = {10.3847/1538-3881/aabc4f},
archivePrefix = {arXiv},
       eprint = {1801.02634},
 primaryClass = {astro-ph.IM},
       adsurl = {https://ui.adsabs.harvard.edu/abs/2018AJ....156..123A}
}

@ARTICLE{astropy:2022,
       author = {{Astropy Collaboration} and {Price-Whelan}, Adrian M. and {Lim}, Pey Lian and {Earl}, Nicholas and {Starkman}, Nathaniel and {Bradley}, Larry and {Shupe}, David L. and {Patil}, Aarya A. and {Corrales}, Lia and {Brasseur}, C.~E. and {N{"o}the}, Maximilian and {Donath}, Axel and {Tollerud}, Erik and {Morris}, Brett M. and {Ginsburg}, Adam and {Vaher}, Eero and {Weaver}, Benjamin A. and {Tocknell}, James and {Jamieson}, William and {van Kerkwijk}, Marten H. and {Robitaille}, Thomas P. and {Merry}, Bruce and {Bachetti}, Matteo and {G{"u}nther}, H. Moritz and {Aldcroft}, Thomas L. and {Alvarado-Montes}, Jaime A. and {Archibald}, Anne M. and {B{'o}di}, Attila and {Bapat}, Shreyas and {Barentsen}, Geert and {Baz{'a}n}, Juanjo and {Biswas}, Manish and {Boquien}, M{'e}d{'e}ric and {Burke}, D.~J. and {Cara}, Daria and {Cara}, Mihai and {Conroy}, Kyle E. and {Conseil}, Simon and {Craig}, Matthew W. and {Cross}, Robert M. and {Cruz}, Kelle L. and {D'Eugenio}, Francesco and {Dencheva}, Nadia and {Devillepoix}, Hadrien A.~R. and {Dietrich}, J{"o}rg P. and {Eigenbrot}, Arthur Davis and {Erben}, Thomas and {Ferreira}, Leonardo and {Foreman-Mackey}, Daniel and {Fox}, Ryan and {Freij}, Nabil and {Garg}, Suyog and {Geda}, Robel and {Glattly}, Lauren and {Gondhalekar}, Yash and {Gordon}, Karl D. and {Grant}, David and {Greenfield}, Perry and {Groener}, Austen M. and {Guest}, Steve and {Gurovich}, Sebastian and {Handberg}, Rasmus and {Hart}, Akeem and {Hatfield-Dodds}, Zac and {Homeier}, Derek and {Hosseinzadeh}, Griffin and {Jenness}, Tim and {Jones}, Craig K. and {Joseph}, Prajwel and {Kalmbach}, J. Bryce and {Karamehmetoglu}, Emir and {Ka{l}uszy{'n}ski}, Miko{l}aj and {Kelley}, Michael S.~P. and {Kern}, Nicholas and {Kerzendorf}, Wolfgang E. and {Koch}, Eric W. and {Kulumani}, Shankar and {Lee}, Antony and {Ly}, Chun and {Ma}, Zhiyuan and {MacBride}, Conor and {Maljaars}, Jakob M. and {Muna}, Demitri and {Murphy}, N.~A. and {Norman}, Henrik and {O'Steen}, Richard and {Oman}, Kyle A. and {Pacifici}, Camilla and {Pascual}, Sergio and {Pascual-Granado}, J. and {Patil}, Rohit R. and {Perren}, Gabriel I. and {Pickering}, Timothy E. and {Rastogi}, Tanuj and {Roulston}, Benjamin R. and {Ryan}, Daniel F. and {Rykoff}, Eli S. and {Sabater}, Jose and {Sakurikar}, Parikshit and {Salgado}, Jes{'u}s and {Sanghi}, Aniket and {Saunders}, Nicholas and {Savchenko}, Volodymyr and {Schwardt}, Ludwig and {Seifert-Eckert}, Michael and {Shih}, Albert Y. and {Jain}, Anany Shrey and {Shukla}, Gyanendra and {Sick}, Jonathan and {Simpson}, Chris and {Singanamalla}, Sudheesh and {Singer}, Leo P. and {Singhal}, Jaladh and {Sinha}, Manodeep and {Sip{H{o}}cz}, Brigitta M. and {Spitler}, Lee R. and {Stansby}, David and {Streicher}, Ole and {{{S}}umak}, Jani and {Swinbank}, John D. and {Taranu}, Dan S. and {Tewary}, Nikita and {Tremblay}, Grant R. and {Val-Borro}, Miguel de and {Van Kooten}, Samuel J. and {Vasovi{'c}}, Zlatan and {Verma}, Shresth and {de Miranda Cardoso}, Jos{'e} Vin{'i}cius and {Williams}, Peter K.~G. and {Wilson}, Tom J. and {Winkel}, Benjamin and {Wood-Vasey}, W.~M. and {Xue}, Rui and {Yoachim}, Peter and {Zhang}, Chen and {Zonca}, Andrea and {Astropy Project Contributors}},
        title = "{The Astropy Project: Sustaining and Growing a Community-oriented Open-source Project and the Latest Major Release (v5.0) of the Core Package}",
      journal = {\apj},
         year = 2022,
        month = aug,
       volume = {935},
       number = {2},
          eid = {167},
        pages = {167},
          doi = {10.3847/1538-4357/ac7c74},
archivePrefix = {arXiv},
       eprint = {2206.14220},
 primaryClass = {astro-ph.IM},
       adsurl = {https://ui.adsabs.harvard.edu/abs/2022ApJ...935..167A}
}

@ARTICLE{dhillon_2022,
       author = {{Dhillon}, V.~S. and {Kennedy}, M.~R. and {Breton}, R.~P. and {Clark}, C.~J. and {Mata S{\'a}nchez}, D. and {Voisin}, G. and {Breedt}, E. and {Brown}, A.~J. and {Dyer}, M.~J. and {Green}, M.~J. and {Kerry}, P. and {Littlefair}, S.~P. and {Marsh}, T.~R. and {Parsons}, S.~G. and {Pelisoli}, I. and {Sahman}, D.~I. and {Wild}, J.~F. and {van Kerkwijk}, M.~H. and {Stappers}, B.~W.},
        title = "{Multicolour optical light curves of the companion star to the millisecond pulsar PSR J2051-0827}",
      journal = {\mnras},
         year = 2022,
        month = oct,
       volume = {516},
       number = {2},
        pages = {2792-2800},
          doi = {10.1093/mnras/stac2357},
archivePrefix = {arXiv},
       eprint = {2208.09249},
 primaryClass = {astro-ph.SR},
       adsurl = {https://ui.adsabs.harvard.edu/abs/2022MNRAS.516.2792D}
}

@ARTICLE{kandel_2023,
       author = {{Kandel}, D. and {Romani}, Roger W.},
        title = "{An Optical Study of the Black Widow Population}",
      journal = {\apj},
         year = 2023,
        month = jan,
       volume = {942},
       number = {1},
          eid = {6},
        pages = {6},
          doi = {10.3847/1538-4357/aca524},
archivePrefix = {arXiv},
       eprint = {2211.16990},
 primaryClass = {astro-ph.HE},
       adsurl = {https://ui.adsabs.harvard.edu/abs/2023ApJ...942....6K}
}

@ARTICLE{van_kerkwijk_2011,
       author = {{van Kerkwijk}, M.~H. and {Breton}, R.~P. and {Kulkarni}, S.~R.},
        title = "{Evidence for a Massive Neutron Star from a Radial-velocity Study of the Companion to the Black-widow Pulsar PSR B1957+20}",
      journal = {\apj},
         year = 2011,
        month = feb,
       volume = {728},
       number = {2},
          eid = {95},
        pages = {95},
          doi = {10.1088/0004-637X/728/2/95},
archivePrefix = {arXiv},
       eprint = {1009.5427},
 primaryClass = {astro-ph.HE},
       adsurl = {https://ui.adsabs.harvard.edu/abs/2011ApJ...728...95V}
}

@ARTICLE{kennedy_2022,
       author = {{Kennedy}, M.~R. and {Breton}, R.~P. and {Clark}, C.~J. and {Mata S{\'a}nchez}, D. and {Voisin}, G. and {Dhillon}, V.~S. and {Halpern}, J.~P. and {Marsh}, T.~R. and {Nieder}, L. and {Ray}, P.~S. and {van Kerkwijk}, M.~H.},
        title = "{Measuring the mass of the black widow PSR J1555-2908}",
      journal = {\mnras},
         year = 2022,
        month = may,
       volume = {512},
       number = {2},
        pages = {3001-3014},
          doi = {10.1093/mnras/stac379},
archivePrefix = {arXiv},
       eprint = {2202.05111},
 primaryClass = {astro-ph.HE},
       adsurl = {https://ui.adsabs.harvard.edu/abs/2022MNRAS.512.3001K}
}

@article{romani_2022,
doi = {10.3847/2041-8213/ac8007},
url = {https://dx.doi.org/10.3847/2041-8213/ac8007},
year = {2022},
month = {jul},
publisher = {The American Astronomical Society},
volume = {934},
number = {2},
pages = {L17},
author = {Romani, Roger W. and Kandel, D. and Filippenko, Alexei V. and Brink, Thomas G. and Zheng, WeiKang},
title = {PSR J0952−0607: The Fastest and Heaviest Known Galactic Neutron Star},
journal = {The Astrophysical Journal Letters}
}

@ARTICLE{simpson_2025,
       author = {{Simpson}, Jordan A. and {Linares}, Manuel and {Casares}, Jorge and {Shahbaz}, Tariq and {Sen}, Bidisha and {Camilo}, Fernando},
        title = "{A GTC spectroscopic study of three spider pulsar companions: line-based temperatures, a new face-on redback, and improved mass constraints}",
      journal = {\mnras},
         year = 2025,
        month = jan,
       volume = {536},
       number = {3},
        pages = {2169-2186},
          doi = {10.1093/mnras/stae2728},
archivePrefix = {arXiv},
       eprint = {2408.11099},
 primaryClass = {astro-ph.HE},
       adsurl = {https://ui.adsabs.harvard.edu/abs/2025MNRAS.536.2169S}
}

@ARTICLE{guo_2022,
       author = {{Guo}, Yunlang and {Wang}, Bo and {Han}, Zhanwen},
        title = "{Formation of black widows through ultracompact X-ray binaries with He star companions}",
      journal = {\mnras},
         year = 2022,
        month = sep,
       volume = {515},
       number = {2},
        pages = {2725-2732},
          doi = {10.1093/mnras/stac1917},
archivePrefix = {arXiv},
       eprint = {2204.06132},
 primaryClass = {astro-ph.HE},
       adsurl = {https://ui.adsabs.harvard.edu/abs/2022MNRAS.515.2725G}
}

@ARTICLE{nelemans_2010,
       author = {{Nelemans}, G. and {Yungelson}, L.~R. and {van der Sluys}, M.~V. and {Tout}, Christopher A.},
        title = "{The chemical composition of donors in AM CVn stars and ultracompact X-ray binaries: observational tests of their formation}",
      journal = {\mnras},
         year = 2010,
        month = jan,
       volume = {401},
       number = {2},
        pages = {1347-1359},
          doi = {10.1111/j.1365-2966.2009.15731.x},
archivePrefix = {arXiv},
       eprint = {0909.3376},
 primaryClass = {astro-ph.SR},
       adsurl = {https://ui.adsabs.harvard.edu/abs/2010MNRAS.401.1347N}
}

@ARTICLE{paczynski_1971,
       author = {{Paczy{\'n}ski}, B.},
        title = "{Evolutionary Processes in Close Binary Systems}",
      journal = {\araa},
         year = 1971,
        month = jan,
       volume = {9},
        pages = {183},
          doi = {10.1146/annurev.aa.09.090171.001151},
       adsurl = {https://ui.adsabs.harvard.edu/abs/1971ARA&A...9..183P}
}
\bibliographystyle{aasjournalv7}

\end{document}